\documentclass[fleqn,usenatbib,useAMS]{mnras}

\DeclareRobustCommand{\VAN}[3]{#2}
\let\VANthebibliography\thebibliography
\def\thebibliography{\DeclareRobustCommand{\VAN}[3]{##3}\VANthebibliography}

\usepackage{amsmath,amsfonts,amssymb} 
\usepackage{bm} 
\usepackage[table,  dvipsnames]{xcolor} 
\usepackage{hyperref} 
\usepackage{nicefrac} 
\usepackage{placeins}

\usepackage[varg]{txfonts} 
\usepackage[T1]{fontenc}

\hypersetup{
	pdftitle = {Migration of low-mass planets in inviscid disks: the effect of cooling on the dynamical corotation torque},
	pdfauthor = {Alexandros Ziampras, Sijme-Jan Paardekooper, Richard P. Nelson},
	pdfsubject = {},
	colorlinks = {true},
	linkcolor = [rgb]{0,0.35,0.7},
	citecolor = [rgb]{0,0.35,0.7},
	filecolor = [rgb]{0.61,0,0},
	urlcolor = [rgb]{0.61,0,0},
}

\usepackage{graphicx}  
\graphicspath{{figures/}}

\newcommand{\DP}[2]{\frac{\partial{#1}}{\partial{#2}}}
\newcommand{\D}[2]{\frac{\text{d}{#1}}{\text{d}{#2}}}

\newcommand{\ap}{a_\mathrm{p}}
\newcommand{\G}{\text{G}}
\newcommand{\Mstar}{M_\star}
\newcommand{\Lstar}{L_\star}
\newcommand{\Rp}{R_\mathrm{p}}
\newcommand{\Mp}{M_\mathrm{p}}

\newcommand{\Hp}{H_\mathrm{p}}

\newcommand{\Mth}{M_\mathrm{th}}

\newcommand{\Rgas}{\mathcal{R}}
\newcommand{\cs}{c_\mathrm{s}}

\newcommand{\OmegaK}{\Omega_\mathrm{K}}

\newcommand{\tauR}{\tau_\mathrm{R}}
\newcommand{\tauP}{\tau_\mathrm{P}}

\newcommand{\taueff}{\tau_\mathrm{eff}}
\newcommand{\kappaR}{\kappa_\mathrm{R}}
\newcommand{\kappaP}{\kappa_\mathrm{P}}
\newcommand{\cv}{c_\mathrm{v}}
\newcommand{\rhomid}{\rho_\mathrm{mid}}
\newcommand{\sigmaSB}{\sigma_\mathrm{SB}}
\newcommand{\vel}{\bm{u}}
\newcommand{\xh}{{x}_\mathrm{h}}
\newcommand{\varpih}{{\varpi}_\mathrm{h}}

\newcommand{\tcool}{t_\mathrm{cool}}
\newcommand{\bcool}{\beta_\mathrm{cool}}
\newcommand{\bsurf}{\beta_\mathrm{surf}}
\newcommand{\bmid}{\beta_\mathrm{mid}}

\newcommand{\Qcool}{Q_\mathrm{cool}}
\newcommand{\Qirr}{Q_\mathrm{irr}}
\newcommand{\Qrad}{Q_\mathrm{rad}}
\newcommand{\Qrelax}{Q_\mathrm{relax}}

\title[Planet migration in radiative disks]{Migration of low-mass planets in inviscid disks:\\the effect of radiation transport on the dynamical corotation torque}

\author[A.~Ziampras et al.]{
Alexandros~Ziampras$^{1}$\thanks{E-mail: a.ziampras@qmul.ac.uk},
Richard~P.~Nelson$^{1}$,
Sijme-Jan~Paardekooper$^{1,2}$
\\
$^{1}$Astronomy Unit, School of Physics and Astronomy, Queen Mary University of London, London E1 4NS, UK\\
$^{2}$Faculty of Aerospace Engineering, Delft University of Technology, Kluyverweg 1, 2600 AA Delft, The Netherlands\\
}

\date{Accepted XXX. Received YYY; in original form ZZZ}
\pubyear{2023}

\begin{document}
\label{firstpage}
\pagerange{\pageref{firstpage}--\pageref{lastpage}}
\maketitle

\begin{abstract}
	Low-mass planets migrate in the type-I regime. In the inviscid limit, the contrast between the vortensity trapped inside the planet's corotating region and the background disk vortensity leads to a dynamical corotation torque, which is thought to slow down inward migration. We investigate the effect of radiative cooling on low-mass planet migration using inviscid 2D hydrodynamical simulations. We find that cooling induces a baroclinic forcing on material U-turning near the planet, resulting in vortensity growth in the corotating region, which in turn weakens the dynamical corotation torque and leads to 2--3$\times$ faster inward migration. This mechanism is most efficient when cooling acts on a timescale similar to the U-turn time of material inside the corotating region, but is nonetheless relevant for a substantial radial range in a typical disk ($R\sim5\text{--}50$\,au). As the planet migrates inwards, the contrast between the vortensity inside and outside the corotating region increases and partially regulates the effect of baroclinic forcing. As a secondary effect, we show that radiative damping can further weaken the vortensity barrier created by the planet's spiral shocks, supporting inward migration. Finally, we highlight that a self-consistent treatment of radiative diffusion as opposed to local cooling is critical in order to avoid overestimating the vortensity growth and the resulting migration rate. 
\end{abstract}



\section{Introduction}
\label{sec:introduction}

More than 5500 exoplanets have been discovered so far\footnote{\url{https://exoplanetarchive.ipac.caltech.edu/}}, and their population shows remarkable diversity in orbital parameters, masses, and multiplicity. The idea that these planets formed in circumstellar disks is supported by both the direct observation of planets embedded in the disk around PDS~70 with VLT \citep{keppler-etal-2018,haffert-etal-2019}, as well as kinematic signatures in observations of gas emission with ALMA \citep[e.g.,][]{teague-etal-2018}. Understanding how planets formed and evolved to their current state is a key goal of modern astrophysics.

The final orbital configuration of a planet is directly affected by its interaction with the disk it is embedded in. The disk exerts a torque on the planet, which causes it to migrate through the disk \citep{goldreich-tremaine-1979,ward-1997a,tanaka-etal-2002}. The migration rate depends on the disk's properties as well as on the planet's mass, with low-mass planets migrating in the rapid type-I regime \citep{ward-1997a}. For a recent review, see \citet{paardekooper-etal-2022}.

Modeling planet--disk interaction over long timescales is not an easy task. The population synthesis approach \citep[e.g.,][]{mordasini-etal-2009a,mordasini-etal-2009b} and N-body modeling \citep[e.g.,][]{coleman-nelson-2014,izidoro-etal-2017} are computationally inexpensive and cover a wide range of parameters in a relatively short time, but rely on analytical prescriptions for the torques acting on the embedded planet(s). Hydrodynamical simulations, on the other hand, are significantly more computationally restrictive, but allow a full treatment of planet--disk interaction that yields more accurate results.

Hydrodynamical modeling is favorable for massive-enough planets, where planet--disk interaction becomes nonlinear \citep{rafikov-2002} and gap opening can occur \citep{crida-etal-2006}. It is also crucial for low-viscosity disks, where the corotation torque can desaturate \citep{paardekooper-etal-2011} and dynamical corotation torques can come into play \citep{mcnally-etal-2017}, as well as for non-isothermal disks, where thermal effects can even reverse the direction of migration \citep{kley-crida-2008,pierens-2015}. In light of the recent paradigm shift towards low-turbulence disks, where accretion is thought to be powered by MHD-driven winds \citep{bai-stone-2013}, and the constraints on the radiative properties of protoplanetary disks through their dust distribution \citep{birnstiel-etal-2018}, the need for understanding how radiative effects can affect planet migration in the nearly inviscid limit is more relevant than ever.

A simplified thermodynamics model that is often employed in simulations of planet--disk interactions is the locally isothermal equation of state, which assumes instant cooling of the gas towards an equilibrium temperature. However, recent models \citep{miranda-rafikov-2019,miranda-rafikov-2020a,ziampras-etal-2020b}, based on observations \citep[e.g.,][]{andrews-etal-2018,oberg-etal-2021}, highlight the importance of a more realistic treatment of thermodynamics and therefore radiative effects, even in regions where cooling occurs on dynamical time scales. Further work has showcased the impact of thermal diffusion \citep{ziampras-etal-2023,miranda-rafikov-2020b} in disk thermodynamics. It therefore becomes clear that an investigation of how radiation transport and in particular thermal diffusion influence planet migration in the marginally optically thin 20--40\,au range is necessary.

In this study, we carry out inviscid radiation hydrodynamics simulations of planet--disk interaction, with the focus being the effect of radiation transport in the type-I regime of planet migration. Our goal is to investigate the source of torques related to radiative effects using a realistic prescription of radiation transport in 2D, extending works that have focused on viscous models \citep[e.g.,][]{kley-crida-2008,pierens-2015} but also providing an explanation on the origin of these ``radiative'' torques and constraining where in the disk they might be important.

We describe our physical and numerical setup in Sect.~\ref{sec:physics-numerics}. We present our results in Sect.~\ref{sec:results}, and discuss them in Sect.~\ref{sec:discussion}. We then summarize our findings in Sect.~\ref{sec:summary}.


\section{Physics and numerics}
\label{sec:physics-numerics}

In this section we lay out our physical and numerical framework. We list the vertically integrated hydrodynamical equations, introduce the sources of cooling used in our models, and describe our numerical setup.

\subsection{Physics}
\label{sub:physics}

We consider a disk of ideal gas with adiabatic index $\gamma=7/5$ and mean molecular weight $\mu=2.35$ around a star with mass $\Mstar$ and luminosity $\Lstar$. The inviscid Navier--Stokes equations then read
\begin{subequations}
	\label{eq:navier-stokes}
	\begin{align}
		\label{eq:navier-stokes-1}
		\DP{\Sigma}{t} + \vel\cdot\nabla\Sigma=-\Sigma\nabla\cdot\vel,
	\end{align}
	\begin{align}
	\label{eq:navier-stokes-2}
		\DP{\vel}{t}+ (\vel\cdot\nabla)\vel=-\frac{1}{\Sigma}\nabla P -\nabla(\Phi_\star+\Phi_\mathrm{p}),
	\end{align}
	\begin{align}
	\label{eq:navier-stokes-3}
		\DP{e}{t} + \vel\cdot\nabla e=-\gamma e\nabla\cdot\vel + Q,
	\end{align}
\end{subequations}
where $\Sigma$, $\vel$ and $P$ denote the surface density, velocity vector, and pressure of the gas, and the internal energy density is given by the ideal gas law as $e=P/(\gamma-1)$. The stellar potential at distance $R$ is $\Phi_\star = -\G\Mstar/R$, where $\G$ is the gravitational constant. We can also define the gas isothermal sound speed $\cs=\sqrt{P/\Sigma}=\sqrt{\Rgas T/\mu}$, where $T$ is the gas temperature and $\Rgas$ is the ideal gas constant. The pressure scale height is then $H=\cs/\OmegaK$, where $\OmegaK=\sqrt{\G\Mstar/R^3}$ is the Keplerian angular velocity, and the disk aspect ratio is $h=H/R$.

The planetary potential follows a Plummer-like prescription with a smoothing length $\epsilon=0.6\Hp$ similar to \citet{mueller-kley-2012} to account for the vertical disk stratification:
\begin{equation}
	\label{eq:planet-potential}
	\Phi_\mathrm{p} = -\frac{\G\Mp}{\sqrt{d^2 + \epsilon^2}},\qquad \bm{d} = \bm{R}-\bm{R}_\mathrm{p}.
\end{equation}

Finally, $Q$ contains any additional radiative terms. In our models the disk is heated by stellar irradiation following the passive, irradiated disk model of \citet{menou-goodman-2004}
\begin{equation}
	\label{eq:Qirr}
	\Qirr = 2\frac{\Lstar}{4\pi R^2}(1-\varepsilon)\frac{\theta}{\taueff},\qquad\theta = R\D{h}{R},
\end{equation}
where $\theta$ is the flaring angle with $\theta=2h/7$ \citep[as $T\propto R^{-3/7}$, see][]{chiang-goldreich-1997}, $\varepsilon$ is the disk albedo (here $\varepsilon=1/2$), and $\taueff$ is an effective optical depth following \citet{hubeny-1990}
\begin{equation}
	\label{eq:taueff}
	\taueff = \frac{3\tauR}{8} + \frac{\sqrt{3}}{4} + \frac{1}{4\tauP}.
\end{equation}
Here, $\tau_{\mathrm{R,P}} = \frac{1}{2}\kappa_\mathrm{R,P}\Sigma$ is the optical depth due to the Rosseland and Planck mean opacities $\kappaR$ and $\kappaP$. We assume $\kappaR=\kappaP=\kappa$, following the opacity model of \citet{lin-papaloizou-1985}. 

We then treat thermal cooling through the disk surfaces as
\begin{equation}
	\label{eq:Qcool}
	\Qcool = -2\frac{\sigmaSB T^4}{\taueff},
\end{equation}
and in-plane radiation transport with a flux-limited diffusion approach \citep[FLD,][]{levermore-pomraning-1981}
\begin{equation}
	\label{eq:Qrad}
	\Qrad = \sqrt{2\pi}H\nabla\cdot\left(\lambda\frac{4\sigmaSB}{\kappaR\rhomid}\nabla T^4\right),\qquad \rhomid = \frac{1}{\sqrt{2\pi}}\frac{\Sigma}{H},
\end{equation}
with the flux limiter $\lambda$ following \citet{kley-1989}. The balance between $\Qirr$, $\Qcool$, and $\Qrad$ results in a temperature profile $T\propto R^{-3/7}$. 

\subsection{Vortensity growth due to cooling}
\label{sub:vortensity-growth}

In this study we often use the vertical component of the gas vortensity, defined in 2D as $\varpi=\nabla\times\vel/\Sigma \,\cdot\hat{z}$, to interpret our results. Taking the curl of Eq.~\eqref{eq:navier-stokes-2} and dividing by $\Sigma$ yields the vortensity equation, which, for a 2D flow with $u_z=0$, reads
\begin{equation}
	\label{eq:vortensity-equation}
	\DP{\varpi}{t} + (\vel\cdot\nabla)\varpi = \frac{\nabla\Sigma\times\nabla P}{\Sigma^3} = \mathcal{S},
\end{equation}
where $\mathcal{S} \propto \nabla\Sigma\times\nabla P$ is a vortensity source term due to baroclinic forcing. For a barotropic flow ($P=P(\Sigma)$), $\varpi$ is conserved along streamlines.

\begin{figure}
	\includegraphics[width=\columnwidth]{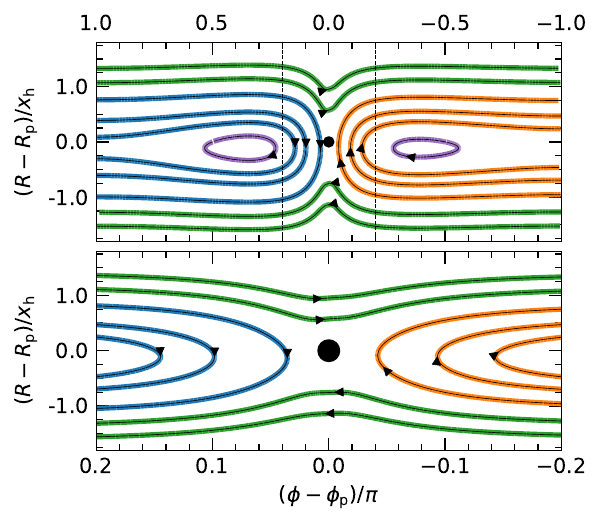}
	\caption{Gas streamlines around an embedded planet (marked with a black dot) in the $\{R,\phi,z=0\}$ plane. Material just outside the planet's corotating region (CR) feels a kick as it shears past the planet (green curves). Inside the CR, material follows closed orbits as it U-turns behind (orange) and ahead (blue) of the planet. Purple orbits mark the L4 and L5 Lagrange points. The bottom panel is a zoom-in of the top between the dashed lines, showing that material U-turning closer to the planet librates closer to the edge of the CR.}
	\label{fig:streamlines}
\end{figure}

Let us now consider a gas parcel in the planet's corotating region. In the planet's corotating frame this parcel follows a streamline with a horseshoe pattern \citep[see e.g.,][]{kley-nelson-2012}, performing a U-turn once behind and once ahead of the planet before completing a closed loop. We can now analyze what happens to the vortensity of the gas parcel during the U-turn at $R=\Rp$, along the radial direction (see schematic in Fig.~\ref{fig:streamlines}).

\subsubsection{Locally isothermal limit}
\label{subsub:isothermal}

In the limit where cooling happens on timescales much shorter than the orbital timescale $\OmegaK^{-1}$, we can assume that $T(R) \propto R^q$ is set by a balance between the radiative terms listed in Sect.~\ref{sub:physics}, and U-turning material is expected to drive vortensity growth for $q\neq0$ \citep{casoli-masset-2009}. The vortensity of gas U-turning behind the planet ($\varpi_\mathrm{b}$) will change according to
\begin{equation}
	\label{eq:vortensity-iso-trailing}
	\mathcal{S}^\mathrm{iso}_\mathrm{b} = \frac{P}{T\Sigma^3} \nabla\Sigma_\mathrm{b}\times\nabla T|_\mathrm{p} = - \frac{P_\mathrm{b}}{T_\mathrm{b}\Sigma_\mathrm{b}^3} \left.\left(\DP{\Sigma}{\phi}\DP{T}{R}\right)\right|_\mathrm{b} = -q\frac{P_\mathrm{b}}{\Rp\Sigma_\mathrm{b}^3}\left.\DP{\Sigma}{\phi}\right|_\mathrm{b},
\end{equation}
where a subscript `p' denotes the value at the planet's radial location. Assuming a dense, circular envelope around the planet such that $\partial_\phi\Sigma|_\mathrm{b}>0$ and $q<0$, this implies that $\mathcal{S}_\mathrm{b}^\mathrm{iso} > 0$. In other words, $\varpi$ will increase as material U-turns behind the planet.
In a similar way, the vortensity of gas U-turning ahead of the planet ($\varpi_\mathrm{a}$) will change as
\begin{equation}
	\label{eq:vortensity-iso-leading}
	\mathcal{S}^\mathrm{iso}_\mathrm{a} = -q\frac{P_\mathrm{a}}{\Rp\Sigma_\mathrm{a}^3}\left.\DP{\Sigma}{\phi}\right|_\mathrm{a},
\end{equation}
resulting in a decrease in $\varpi$ since $\partial_\phi\Sigma|_\mathrm{a}<0$.

Finally, assuming that the U-turns are symmetric such that $\partial_\phi\Sigma|_\mathrm{b} = -\partial_\phi\Sigma|_\mathrm{a}$ and $x_\mathrm{a} = x_\mathrm{b}$ for $x\in[P,\Sigma,T]$, we have that $\mathcal{S}^\mathrm{iso}_\mathrm{a} + \mathcal{S}^\mathrm{iso}_\mathrm{b} = 0$. In other words, while $\varpi$ will increase (decrease) as material U-turns behind (ahead of) the planet, the total change along a closed horseshoe loop is zero and $\varpi$ is ultimately conserved.

\subsubsection{Adiabatic limit}
\label{subsub:adiabatic}

When cooling is inefficient, the parcel contracts (expands) adiabatically as it approaches (recedes from) the planet. This process is reversible and entropy is conserved along the streamline. Each streamline therefore retains its own entropy. Defining the entropy function $K=P/\Sigma^\gamma$, with $\Sigma\propto R^s$, we initially have that $K(R)\propto R^\xi$, with $\xi=q + (1-\gamma)s$. Assuming, without loss of generality, that $\xi<0$, a high-entropy streamline at $R<\Rp$ U-turning behind the planet will advect into a low-entropy radial zone at $R>\Rp$, resulting in an entropy discontinuity near $R\approx\Rp$ and a nonzero $\mathcal{S}_\mathrm{b}$. The low-entropy streamline will, of course, also U-turn ahead of the planet into the high-entropy zone at $R<\Rp$, resulting in $\mathcal{S}_\mathrm{a} = -\mathcal{S}_\mathrm{b}$ and a net $\mathcal{S}^\mathrm{adb}=0$. Furthermore, after a few libration timescales (see Eq.~\eqref{eq:tlib}), entropy will be completely mixed in the horseshoe region and therefore
\begin{equation}
	\mathcal{S}^\mathrm{adb} \propto \nabla\Sigma\times\nabla P \propto \nabla\Sigma\times\nabla K = 0.
\end{equation}
In our models, $\xi\approx-0.03$ and therefore vortensity growth in adiabatic models is negligible. We do, however, explore different values of $\xi$ in Appendix~\ref{apdx:adiabatic}.

\subsubsection{Finite cooling timescale}
\label{subsub:cooling}

When cooling acts on a finite timescale, this calculation becomes complicated. We can however show that the two source terms ahead of and behind the planet ($\mathcal{S}_\mathrm{a}$, $\mathcal{S}_\mathrm{b}$) are asymmetric such that vortensity always increases along a pair of streamlines U-turning in opposite directions. We provide a non-rigorous derivation in Appendix~\ref{apdx:vortensity-math} as a proof of concept.

\subsection{Dynamical corotation torque}
\label{sub:dct}

In a sufficiently laminar disk, where mixing between the planet's horseshoe region and the background disk is negligible, a migrating planet will experience a dynamical corotation torque \citep{mcnally-etal-2017}
\begin{equation}
	\label{eq:dct}
	\Gamma_\mathrm{h} = 2\pi \left(1-\frac{\varpi(\Rp)}{\varpi_\mathrm{h}}\right) \Sigma_\mathrm{p} \Rp^2 \xh \Omega_\mathrm{p}\left(\D{\Rp}{t} - u_R\right),
\end{equation}
where the subscript `p' denotes quantities at the planet's radial position and $\xh$ is an estimate of the half-width of the corotating region in the 2D approximation following \citet{paardekooper-etal-2010}
\begin{equation}
	\label{eq:corotation-width}
	\xh = \frac{1.1}{\gamma^{1/4}}\left(\frac{0.4}{\epsilon/H}\right)^{1/4}\sqrt{\frac{\Mp}{h\Mstar}}\Rp.
\end{equation}
In Eq.~\eqref{eq:dct}, $\varpi(\Rp)$ denotes the vortensity of the background disk at $\Rp$ and $\varpih$ refers to the characteristic vortensity enclosed in the planet's horseshoe region, which would correspond to the vortensity of the initial radial location of the planet ($R_0$) if $\varpi$ is conserved inside the corotation region.

Provided that the background disk radial velocity $u_R$ is small compared to the planet's migration speed, this dynamical corotation torque is positive for a planet migrating inwards and becomes stronger as $\varpi(\Rp)/\varpih$ increases (assuming $\Sigma\propto R^{-3/2}$ or shallower), eventually stalling the planet \citep{paardekooper-2014}.
Based on the discussion in Sects.~\ref{subsub:adiabatic}--\ref{subsub:cooling}, a vortensity source term in the horseshoe region would cause $\varpih$ to evolve as $\varpih(t)\approx \varpi(R_0) + \int_{0}^{t}\mathcal{S}^\mathrm{rad}(t^\prime)\,\mathrm{d}t^\prime$, weakening the dynamical corotation torque and speeding up inward migration, possibly preventing the planet from stalling.

\subsection{Numerics}
\label{sub:numerics}

We solve the equations Eq.~\eqref{eq:navier-stokes} using the Godunov-scheme numerical hydrodynamics code \texttt{PLUTO} \texttt{v.4.4} \citep{mignone-etal-2007}. We use a cylindrical grid $\{R,\phi\}$ with $N_R\times N_\phi = 805\times 3141$ cells, logarithmically spaced in $R$ and uniformly spaced in $\phi$. Our grid covers the radial extent $R\in[0.4,2]\,R_0$ with $R_0=30$\,au, and $\phi\in[0,2\pi]$. 

Our initial conditions assume power-law profiles for $\Sigma$ and $T$ with
\begin{equation}
	\label{eq:initial-conditions}
	\Sigma_0 = \Sigma_\mathrm{ref} \left(\frac{R}{R_0}\right)^{-1},\qquad
	T_0 = T_\mathrm{ref} \left(\frac{R}{R_0}\right)^{-3/7},
\end{equation}
with $\Sigma_\mathrm{ref} = 56.7\,\text{g}/\text{cm}^2$ and $T_\mathrm{ref} = 21\,\text{K}$, or equivalently $h_0=0.05$, obtained by a balance between $\Qirr$ and $\Qcool$ (see Eqs.~\eqref{eq:Qirr}~\&~\eqref{eq:Qcool}). With these parameters, our chosen resolution yields 25 cells per scale height at $R=R_0$. The disk surface density corresponds to $1700\,\text{g}/\text{cm}^2$ at 1\,au. The chosen set of parameters also results in a cooling timescale $t_\text{cool}\approx1.6\,\OmegaK^{-1}$ at $R=R_0$ \citep[][see also Eqs.~\eqref{eq:bcool}~\&~\eqref{eq:bsurf-bmid}]{ziampras-etal-2023}. 

The velocity field is initialized with a Keplerian profile corrected for pressure support, and the planet is placed at $\Rp=R_0$ with $\Mp=2\times10^{-5}\,\Mstar \approx 6.7\,M_\oplus$. The disk boundaries are periodic in azimuth and closed in the radial direction, and we apply the wave-damping prescription of \citet{devalborro-etal-2006} in the radial zones $R<0.525$ and $R>1.525$ with a damping timescale of 0.1 orbits at the respective boundary.

The planet is integrated using an adaptation of the N-body module by \citet{thun-kley-2018}, where the correction due to the neglect of disk self-gravity by \citet{baruteau-masset-2008b} is applied to models where the planet is allowed to migrate. The indirect term by the planet and star orbiting their common center of mass is included in all models, but the star feels the gravitational influence of the disk only in models where the planet can migrate \citep{crida-etal-2022}. The planet grows to its final mass over 10 orbits using the formula in \citet{devalborro-etal-2006}.

In all runs we use the FARGO algorithm \citep{masset-2000}, implemented in \texttt{PLUTO} by \citet{mignone-etal-2012}. We use the \texttt{hllc} Riemann solver \citep{toro-etal-1994} with the \texttt{VAN\_LEER} limiter \citep{vanleer-1974}, 2nd-order \texttt{RK2} timestepping and the 3rd-order \texttt{WENO3} reconstruction \citep{yamaleev-carpenter-2009}.

We run two sets of models, where the planet is either fixed or allowed to migrate through the disk. In each set we carry out the following models:
\begin{description}
	\item[\emph{radiative}:] realistic disk with $Q = \Qirr + \Qcool + \Qrad$,
	\item[\emph{adiabatic}:] no radiative terms in Eq.~\eqref{eq:navier-stokes-3} (i.e., $Q=0$),
	\item[\emph{locally isothermal}:] Eq.~\eqref{eq:navier-stokes-3} is not solved; $T=T_0(R)$.
\end{description}
These models are sometimes tagged ``\texttt{rad}'', ``\texttt{adb}'', and ``\texttt{iso}'' in our plots. For consistency, we always use orange, blue, and green respectively to refer to each model in all plots.

\section{Results}
\label{sec:results}

In this section we present the results of our numerical models. We first focus on the vortensity evolution in models with fixed planets, before moving on to models where the planet is allowed to migrate. We then investigate how different cooling regimes affect the vortensity evolution.

\subsection{Fixed planet}
\label{sub:fiducial-static}

\begin{figure*}
	\includegraphics[width=\textwidth]{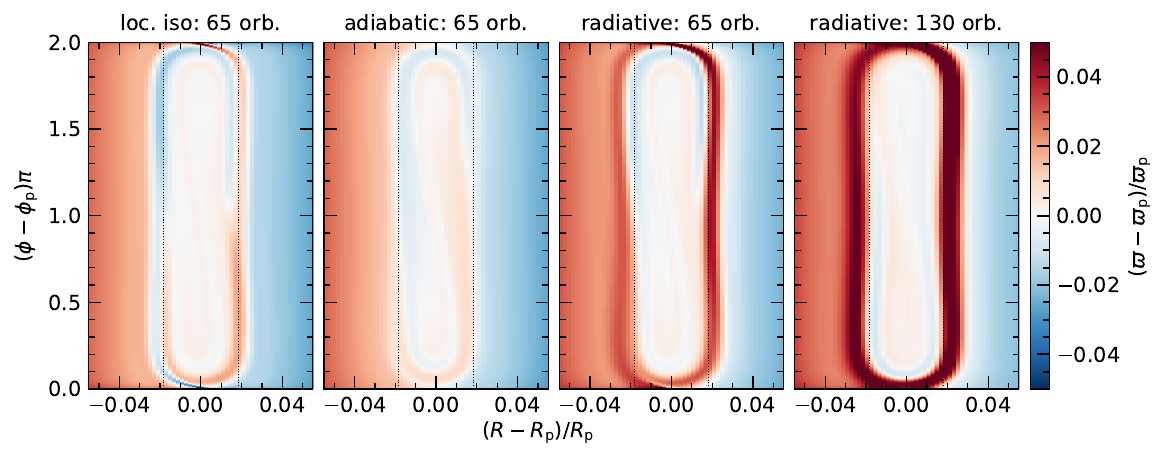}
	\caption{Two-dimensional heatmaps showing the vortensity $\varpi$ in the planet's corotating region (CR). The first three panels from left to right compare different models at $t=\tau_\text{lib}$ (see Eq.~\eqref{eq:tlib}), showing that $\varpi$ is being mixed throughout in the CR but an excess develops in the radiative run. The rightmost panel shows that the excess grows with time in the radiative model. Vertical dashed lines mark the edges of the CR at $\pm\xh$ (see Eq.~\eqref{eq:corotation-width}).}
	\label{fig:fiducial-vortensity}
\end{figure*}

We first compare the vortensity evolution in our models with a fixed planet. Even though there is initially a radial vortensity gradient through the disk ($\varpi_0(R)\approx\OmegaK/2$), the vortensity is mixed in the planet's corotating region. This process happens over the libration timescale
\begin{equation}
	\label{eq:tlib}
	\tau_\text{lib} = \frac{8\pi\Rp}{3\Omega_\text{p}\xh}.
\end{equation}
After a few $\tau_\text{lib}$, $\varpi$ is flat inside the corotating region and the corotation torque vanishes.

\begin{figure}
	\includegraphics[width=\columnwidth]{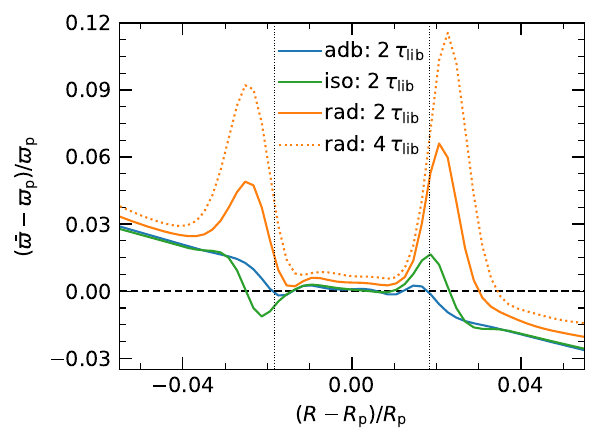}
	\caption{Azimuthally averaged perturbed vortensity in the planet's corotating region (CR) for the models in Fig.~\ref{fig:fiducial-vortensity}. The vortensity excess in the radiative model grows with time and affects the entire CR.}
	\label{fig:fiducial-vortensity-mean}
\end{figure}

In Fig.~\ref{fig:fiducial-vortensity} we compare the vortensity evolution in our fiducial models with a fixed planet after $1\,\tau_\text{lib}$. In all models $\varpi$ undergoes phase mixing and is roughly constant in the corotating region after a few libration timescales. In the radiative model, however, a vortensity excess is visible (in red) at $R\approx\Rp\pm\xh$, indicating that vortensity is not conserved. We plot the azimuthally averaged vortensity in Fig.~\ref{fig:fiducial-vortensity-mean}, where we show that the excess grows with time and affects the vortensity profile in the entire corotating region.

\begin{figure}
	\includegraphics[width=\columnwidth]{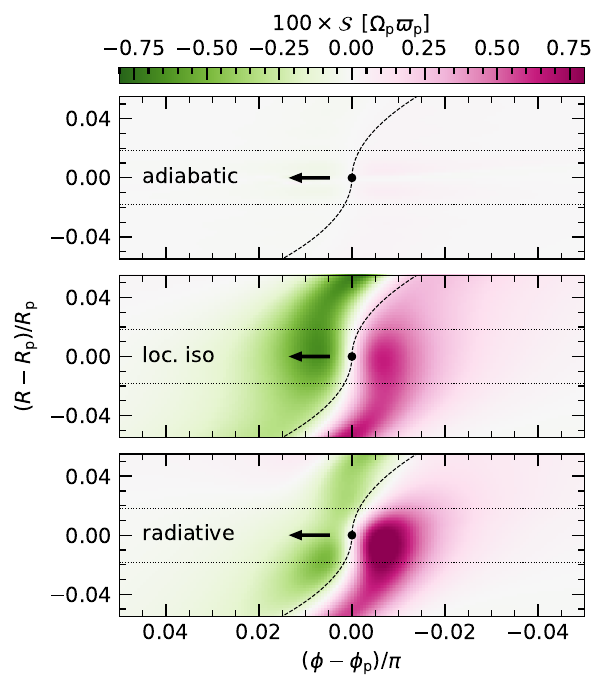}
	\caption{The baroclinic forcing term $\mathcal{S}$ in Eq.~\eqref{eq:vortensity-equation}, which leads to vortensity growth. This term is zero throughout the corotating region in the adiabatic model, averages to zero in the locally isothermal model, but is asymmetric in the radiative model such that vortensity grows over time. Black curves mark the position of the planetary spirals following \citet{rafikov-2002}.}
	\label{fig:fiducial-source}
\end{figure}

This excess, driven by cooling and unrelated to vortensity growth due to the Rossby Wave Instability \citep{lovelace-1999}, is due to the source term $\mathcal{S}$ in Eq.~\eqref{eq:vortensity-equation}, which acts to increase $\varpi$ as material U-turns near the planet (see Sect.~\ref{sub:vortensity-growth}). To verify this, we plot $\mathcal{S}$ for all models at $t=\tau_\text{lib}$ in Fig.~\ref{fig:fiducial-source}. Our findings agree with the expectations in Sect.~\ref{sub:vortensity-growth}: $\mathcal{S}$ is zero in the adiabatic model, nonzero but antisymmetric about the planet in the locally isothermal model such that it averages to zero, and asymmetric in the radiative model such that $\mathcal{S} > 0$ in the corotating region. As a result, $\varpi$ in the corotating region only increases in models with cooling, and in particular on streamlines that U-turn very close to the planet. Since these streamlines librate at the edges of the corotating region, the excess piles up at $R\approx\Rp\pm\xh$ (see Figs.~\ref{fig:fiducial-vortensity}, \ref{fig:fiducial-vortensity-mean}).

Given that the dynamical corotation torque for an inwardly migrating planet is proportional to $\varpi_\text{p}/\varpi_\text{h}-1$, where $\varpi_\text{p}$ is the background disk vortensity at the planet's location and $\varpi_\text{h}$ is the characteristic vortensity trapped in the corotating region \citep{mcnally-etal-2017}, the vortensity growth we found in our radiative models should result in faster inward migration. We investigate this in Sect.~\ref{sub:fiducial-migrating}.

\subsection{Migrating planet}
\label{sub:fiducial-migrating}

\begin{figure}
	\includegraphics[width=\columnwidth]{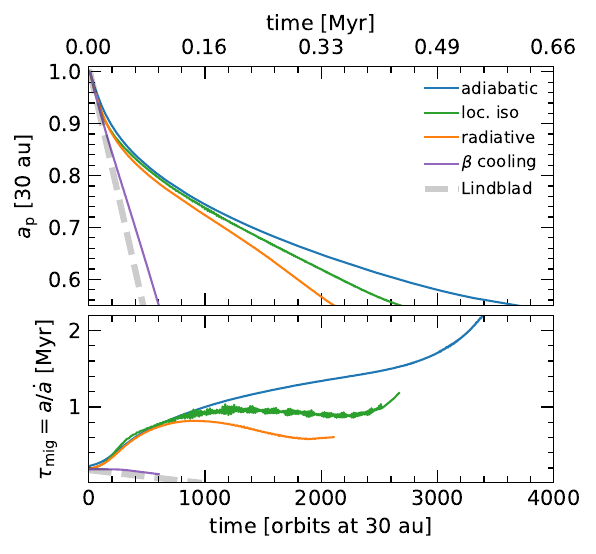}
	\caption{Migration tracks for the models with migrating planets. The radiative model (orange) initially migrates at the locally isothermal rate (green) for $t\lesssim200\,P_0$, then matches the adiabatic rate (blue) until $t\sim600\,P_0$, and afterwards diverges significantly as the planet speeds up. This is also evident in the migration timescales $\tau_\text{mig}$ (bottom), which match between the radiative and adiabatic models for $t\sim200$--600\,$P_0$, but then show that the planet migrates roughly 2--3$\times$ as fast in the radiative model compared to the adiabatic. The dashed curve marks the migration rate without corotation torques. The purple curve (discussed in detail in Sect.~\ref{sub:local-cooling}) shows a model where thermal diffusion was omitted, resulting in unphysically rapid migration at the rate given by Lindblad torques only. Given that we use a wave damping zone for $R<0.53\,R_0$, we exclude data for $\ap<0.55\,R_0$. The damping zone interface is also the reason why migration slows down in all models for $\ap\lesssim0.6\,R_0$}.
	\label{fig:migration-tmig}
\end{figure}

We now repeat the models in Sect.~\ref{sub:fiducial-static}, allowing the planet to migrate after 10 orbits. The migration tracks for these models are shown in Fig.~\ref{fig:migration-tmig}.

Initially, the planet migrates inwards faster in the radiative model than in the adiabatic. This can be attributed to a combination of a weaker dynamical corotation torque due to the vortensity growth mechanism we described in Sects.~\ref{sub:vortensity-growth}~\&~\ref{sub:fiducial-static}, as well as an effectively smaller $\gamma$ due to cooling \citep{paardekooper-etal-2010}.

After 200 orbits, migration in the radiative model slows down to match the adiabatic model. This becomes clear when comparing the migration timescales $\tau_\text{mig}=a/\dot{a}$, which become equal for the radiative and adiabatic models for $t\sim200$--600 orbits. This can be interpreted as the result of a combination of effects. For one, the dynamical corotation torque scales with the planet's migration speed, see Eq.~\eqref{eq:dct}, therefore partially slowing the planet down. In addition, the vortensity excess via baroclinic forcing becomes less important as the planet migrates inwards, because the contrast to the local vortensity ($\varpi_\text{p}/\varpi_\text{h}$) increases faster than the vortensity input via baroclinic forcing in the corotating region when the planet migrates quickly.

\begin{figure}
	\includegraphics[width=\columnwidth]{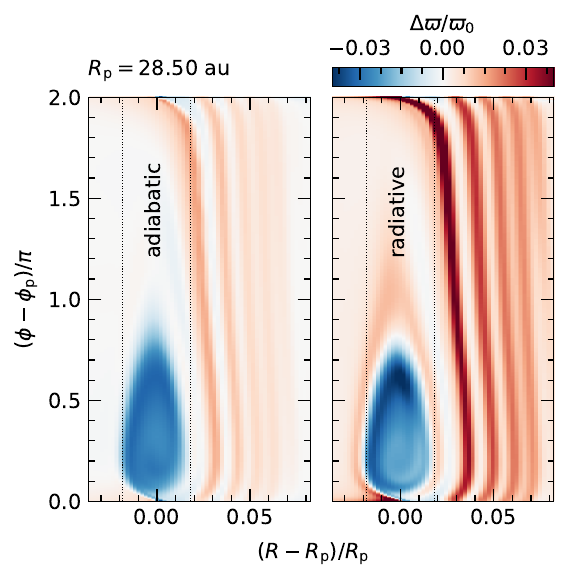}
	\caption{Similar to Fig.~\ref{fig:fiducial-vortensity}, but for the models with migrating planets. The corotating region is now tadpole-shaped, with a vortensity excess around it (resembling a flame) in the radiative model. The vortensity-rich streamlines U-turning behind the planet are no longer confined to the corotating region, as the planet has drifted inwards by the time they would have reached the planet from the front.}
	\label{fig:migrating-vortensity}
\end{figure}

To support this argument we show that vortensity growth is not as efficient in the case of a migrating planet. In part, this happens because the mechanism relies on material U-turning near the planet repeatedly. As the planet migrates inwards, however, material U-turning behind the planet is no longer trapped in the corotating region. Instead, the latter is now a tadpole-shaped zone in front of the planet \citep{masset-papaloizou-2003,papaloizou-etal-2007}, and the vortensity-rich streamlines behind the planet are simply left behind the planet's wake, unable to sustain the feedback loop that led to substantial vortensity growth in the fixed case in Sect.~\ref{sub:fiducial-static}. We illustrate this picture in Fig.~\ref{fig:migrating-vortensity}.

However, after $t\sim600$~orbits, the planet speeds up in the radiative model, migrating roughly $3\times$ and $2\times$ faster compared to the adiabatic and locally isothermal models, respectively. We interpret this as a combination of two effects: a rapidly migrating planet will not capture as many U-turning streamlines, weakening vortensity growth due to cooling. This results in the planet slowing down over time as the vortensity contrast to the background disk increases and the dynamical corotation torque becomes more positive ($\Gamma_\mathrm{h}\propto \text{d}\Rp/\text{d}t$), which is why the radiative model briefly matches the adiabatic migration timescale at $t\sim200$~orbits. By slowing down, however, the planet can capture more vortensity-rich U-turning streamlines, enhancing $\varpih$ and weakening the dynamical corotation torque, causing the planet to speed up once again. The two effects seemingly reach a balance after $\sim1000$ orbits, with the planet migrating inwards without slowing down. We nevertheless stress that this is only a qualitative explanation, and as such we cannot rule out the possibility of runaway migration in the radiative model \citep[e.g.,][]{pierens-2015}.

\subsection{Different cooling timescales}
\label{sub:beta-dependency}

In the previous sections we showed how cooling can speed up inward migration. We also showed that in the limit of both inefficient and rapid cooling (adiabatic and isothermal, respectively), this effect is absent. We therefore expect that the effect depends on the cooling timescale $\bcool=\tcool\OmegaK$, and is maximized for a critical value.

To measure this dependency of vortensity growth on $\bcool$, we run a set of radiative models where we vary the reference surface density $\Sigma_\text{ref}$ to control $\bcool$. For these models, the planet is kept fixed at $R_0$. We compute the cooling timescale following the prescription in \citet{miranda-rafikov-2020b}
\begin{equation}
	\label{eq:bcool}
	\bcool^{-1} = \bsurf^{-1} + \bmid^{-1},
\end{equation}
where $\bsurf$ and $\bmid$ are the cooling timescales for surface and in-plane cooling and are given by \citep{ziampras-etal-2023}
\begin{equation}
	\label{eq:bsurf-bmid}
	\begin{split}
		\bsurf&=\frac{\Sigma\cv T}{|\Qcool|}\OmegaK,\quad \cv=\frac{\Rgas}{\mu(\gamma-1)}\\
		\bmid&=\frac{\OmegaK}{\eta}\left(H^2+\frac{l_\text{rad}^2}{3}\right),\quad \eta=\frac{16\sigmaSB T^3}{3\kappaR\rhomid^2\cv}, \quad l_\text{rad}=\frac{1}{\kappaP\rhomid}.
	\end{split}
\end{equation}

\begin{figure}
	\includegraphics[width=\columnwidth]{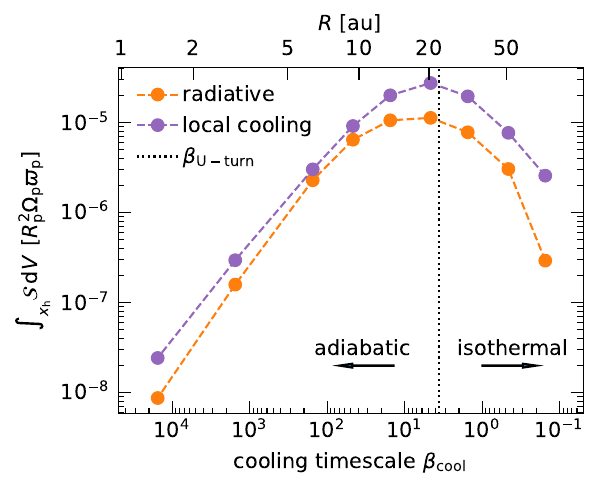}
	\caption{The baroclinic source term $\mathcal{S}$ in Eq.~\eqref{eq:vortensity-equation}, integrated over the corotating region, as a function of the cooling timescale $\bcool$ or the corresponding distance $R$ for our models. The function peaks where $\bcool\approx\beta_\text{U-turn}$ (vertical line), or $R\sim10\text{--}35$\,au for our setup. The orange curve shows the results for our radiative models, while the purple curve shows models without thermal diffusion (discussed in Sect.~\ref{sub:local-cooling}).}
	\label{fig:source-rad}
\end{figure}

We then compute the quantity $\mathcal{S}$ similar to Fig.~\ref{fig:fiducial-source} after 20 planetary orbits and integrate it over the corotating region for each model to obtain a proxy for the vortensity growth rate. We plot the results in Fig.~\ref{fig:source-rad} (orange curve), where we show that this integral peaks at $\bcool\approx 4.3$ and approaches zero in the adiabatic and locally isothermal limits.

The peak agrees very well with the estimate of the U-turn timescale $\beta_\text{U-turn} \approx h \tau_\text{lib} \OmegaK\approx3.65$ in \citet{baruteau-masset-2008a}, highlighted with a vertical line in Fig.~\ref{fig:source-rad}. This suggests that the vortensity growth rate is maximized when the cooling timescale is comparable to the U-turn timescale of material in the corotating region. In other words, the effect of cooling is strongest when it happens over a timescale similar to the time over which U-turning material interacts dynamically with the planet.

\subsection{The effect of thermal diffusion}
\label{sub:local-cooling}

Our radiative models discussed above include a treatment of thermal diffusion along the disk plane through the flux-limited diffusion (FLD) approximation. Thermal diffusion serves to smooth out temperature gradients in the disk, and therefore could affect the vortensity growth rate.

A more simplistic (but less accurate) approach would be to implement radiation transport as a local cooling term in the energy equation, similar to the method in \citet{miranda-rafikov-2020b} \citep[see][for a comparison]{ziampras-etal-2023}. This approach, however, misses the diffusive component of in-plane radiation transport. To investigate the difference, we run a set of models where we replace $\Qirr+\Qcool+\Qrad$ in Eq.~\eqref{eq:Qrad} with a local cooling term $\Qrelax$ given by \citep[e.g.,][]{gammie-2001}
\begin{equation}
	\label{eq:Qrelax}
	\Qrelax = -4\Sigma\cv\frac{T-T_0}{\beta}\OmegaK,
\end{equation}
where $\cv$ is the heat capacity at constant volume and $\beta=\bcool$ is the cooling timescale for both surface and in-plane cooling (see Eqs.~\eqref{eq:bcool}~\&~\eqref{eq:bsurf-bmid}). The factor of 4 is a correction following \citet{ziampras-etal-2023}, obtained by linearizing $\Qcool$ in Eq.~\eqref{eq:Qcool}. We note that \citet{dullemond-etal-2022} cite a correction of $4+b$ with $b = \D{\log\kappa}{\log T} \approx 2$ instead, but this applies only in the optically thin limit where $\taueff\propto \tau^{-1}\propto\kappa^{-1}$ (see Eq.~\eqref{eq:taueff}). In the optically thick limit one would instead use $4-b$ since $\taueff\propto \tau$, but since we have $\tau\sim2 \Rightarrow \taueff\sim 1$ at $R=R_0$ in our models we ignore $b$.

In Fig.~\ref{fig:migration-tmig} we also plot the migration tracks for a run with local, $\beta$ cooling. We find that the planet migrates much faster when thermal diffusion is omitted, leading to drastically faster migration. In addition, we compute the integrated baroclinic term as a function of $\bcool$ similar to Sect.~\ref{sub:beta-dependency} and include it in Fig.~\ref{fig:source-rad} (purple curve). While the function also peaks at $\bcool\approx\beta_\text{U-turn}$, it consistently overestimates the vortensity growth rate compared to our radiative models where thermal diffusion is included. This suggests that, while cooling results in vortensity growth due to baroclinic forcing and therefore faster inward migration, thermal diffusion acts to partially suppress this effect.

\subsection{Radiative damping of spiral shocks}
\label{sub:secondary-effects}

While the focus of this study has been how radiative cooling can induce vortensity growth via baroclinic effects, it is worth highlighting that an additional source of vortensity exists in the planet's vicinity. The planet's spiral arms will steepen into shocks as they propagate through the disk \citep{rafikov-2002}, resulting in a vortensity excess \citep{lin-papaloizou-2010}. This creates a vortensity pileup about $\pm x_\mathrm{sh}$ away from the planet, where $x_\mathrm{sh}$ is the shock distance following \citet{zhu-etal-2015} \citep[see also][]{goodman-rafikov-2001}
\begin{equation}
	\label{eq:xshock}
	x_\mathrm{sh} \approx 0.93\left(\frac{\gamma+1}{12/5}\frac{\Mp}{\Mth}\right)^{-2/5}H,\qquad \Mth=\frac{2\cs^3}{3\G\Omega_\mathrm{p}}\approx h_\mathrm{p}^3 \Mstar.
\end{equation}
Here, $\Mth$ is the thermal mass at the location of the planet \citep{goodman-rafikov-2001}.

However, radiative cooling is expected to weaken the spiral shocks \citep{miranda-rafikov-2020a,zhang-zhu-2020,ziampras-etal-2023}, possibly resulting in weaker vortensity growth. This would imply that, while a planet in an adiabatic or isothermal model would also need to overcome the dynamical corotation torque due to the vortensity ``barrier'' created by its own spiral shocks, this effect is mitigated or even absent in a radiative model. The result would be faster inward migration when $\beta\sim1$, where the damping effect of radiative cooling on spiral shocks is maximized \citep{miranda-rafikov-2020a,ziampras-etal-2023}. This condition is met in our models, as we have $\beta\sim1.5$ at $R=30$\,au (see Fig.~\ref{fig:source-rad}).

\begin{figure}
	\includegraphics[width=\columnwidth]{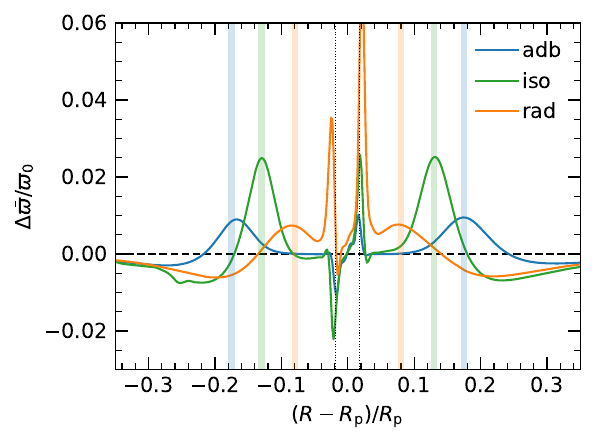}
	\caption{Azimuthally averaged perturbed vortensity for static planet models at $t=2\,\tau_\mathrm{lib}$, similar to Fig.~\ref{fig:fiducial-vortensity-mean} but normalized to the Keplerian vortensity. Colored bands mark the location where spiral shocks deposit vortensity in the disk. The radiative model shows a weaker vortensity excess due to spiral shocks, closer to the planet.}
	\label{fig:fiducial-vortensity-mean-shock}
\end{figure}
\begin{figure}
	\includegraphics[width=\columnwidth]{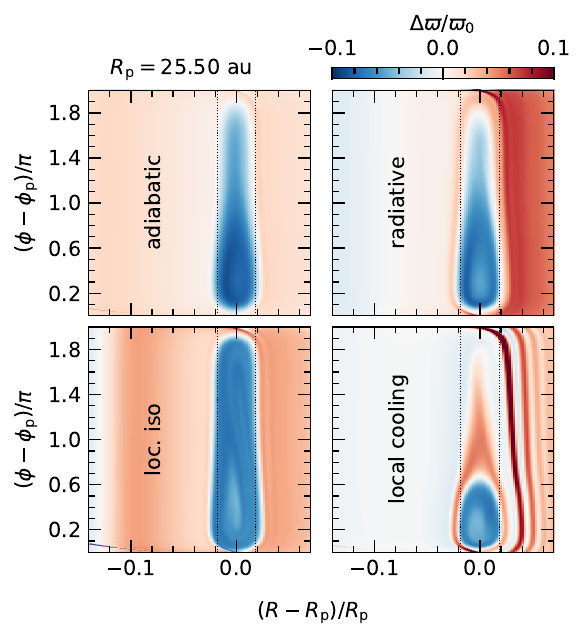}
	\caption{Similar to Fig.~\ref{fig:migrating-vortensity}, with a focus on the vortensity profile ahead of (inside) the planet's orbit. Models with radiative cooling (right) show practically no vortensity excess due to spiral shocks.}
	\label{fig:migrating-vortensity-shock}
\end{figure}
Figure~\ref{fig:fiducial-vortensity-mean-shock} shows a comparison of the azimuthally averaged perturbed vortensity for our fiducial models similar to Fig.~\ref{fig:fiducial-vortensity-mean}, but covering a wider radial range and normalized to the Keplerian vortensity to highlight any excess. Here, it becomes clear that radiative cooling interferes with the damping of spiral shocks, resulting in a vortensity excess that is weaker and closer to the planet. Finally, in Fig.~\ref{fig:migrating-vortensity-shock} we show a heatmap of the perturbed vortensity for models with migrating planets, where we find that the excess due to spiral shocks is practically absent in models with radiative cooling.

We expect that this effect is of secondary importance compared to the baroclinic forcing we discuss in this paper. Nevertheless, we stress that this can further enhance the effect of radiative cooling on inward migration such that baroclinic forcing in the horseshoe region ($\beta\sim\beta_\text{U-turn}\sim10$) and radiative damping of shocks ($\beta\sim1$) can operate efficiently together for $\beta\sim1\text{--}10$. Such an investigation is beyond the scope of this paper, but could be the subject of future work.

\section{Discussion}
\label{sec:discussion}

In this section we discuss our findings in the context of previous studies and more realistic physics. We also highlight the relevance of our results in planet migration.

\subsection{Previous radiative models}

\citet{lega-etal-2014} showed that radiative cooling results in the ``cold finger'' effect in the corotating region, as material is colder and denser after a U-turn compared to an adiabatic case. This would formally result in vortensity generation due to the effect discussed in this paper, but only if the cooling timescale is right (see Fig.~\ref{fig:source-rad}). \citet{pierens-2015} also investigated the effect of radiation transport on planet migration, but did not report on the mechanism described here. \citet{yun-etal-2022} similarly carried out 3D radiative hydrodynamic simulations of planet migration, showcasing the ``cold finger'' effect but not finding evidence of cooling-induced vortensity growth.

This is likely due to the fact that all of these studies probed the inner 1--5\,au, where the cooling timescale is $\bcool\sim10^3\text{--}10^4\gg\beta_\text{U-turn}$, such that the effect we describe was too inefficient (see Fig.~\ref{fig:source-rad}). For a solar-type star, we expect cooling-induced vortensity growth to be most efficient at $R\sim10\text{--}35$\,au (see Fig.~\ref{fig:source-rad}).

\subsection{More realistic physics}

In this paper we aimed to isolate the effect of radiative cooling on the migration rate of low-mass planets. In reality, however, many more processes that we did not include and which can directly or indirectly affect migration could come into play.

For example, accretion heating can give rise to thermal torques \citep[][who also included thermal diffusion]{benitez-etal-2015,masset-2017} or even result in vortex formation \citep{cummins-etal-2022} that could affect migration \citep{lega-etal-2021}. The Hall effect in non-ideal MHD can lead to a torque in the disk midplane, which strongly affects migration \citep{mcnally-etal-2017, mcnally-etal-2018}. In addition, the torque by an MHD-driven wind could reverse the direction of migration \citep{kimmig-etal-2020}. Finally, including the vertical direction can give rise to buoyancy-related torques \citep{zhu-etal-2012,mcnally-etal-2020} as well as introducing a vertical dependence of the MHD wind-driven torques \citep{wafflard-fernandez-lesur-2023}.

We expect that a more realistic picture should include all such effects, but also note that a treatment of cooling---and in particular radiative diffusion---would likely regulate their contribution to the total torque. This could be an interesting topic for future work.

\subsection{Sustaining, retriggering, or exaggerating the effect}

While the dynamical corotation torque will not fully halt inward migration, it can slow the planet down enough for it to open a partial gap and slowly transition into the much slower type-II migration \citep{ward-1997b}. Alternatively, migration can halt as the planet approaches the inner edge of its gap and the (negative) outer Lindblad torque is suppressed \citep[though see][for a discussion on the effect of vortices]{mcnally-etal-2019a}.

In both cases the corotating region will span the full disk azimuth, trapping trailing vortensity-rich streamlines and retriggering rapid inward migration if the planet can break the stall \citep[e.g.,][]{mcnally-etal-2019a}. This could lead to a sustained faster inward migration, or a series of migration stalls and restarts. Investigating this scenario could be the subject of future work.

In the case of a more traditionally viscous disk, one or more zero-torque regions (``migration traps'') could exist in the disk, such as near icelines where the background disk density and temperature profiles change \citep[e.g.,][]{coleman-nelson-2016}. In such a scenario, the (static, outward) corotation torque remains unsaturated and balances against the (inward) Lindblad torque, allowing the planet to remain stationary. Here, the cooling-induced vortensity growth we discuss could result in the planet migrating either inwards or outwards (depending on the accretion velocity of the disk, see Eq.~\eqref{eq:dct}) at a rate that will strongly depend on the disk properties \citep{pierens-2015}.

Our analysis in Sect.~\ref{sub:local-cooling} further showed that thermal diffusion plays a critical role in the vortensity generation and subsequent torque by smoothing temperature gradients. It is therefore important to treat radiation transport appropriately (using e.g., FLD instead of a local cooling prescription for the in-plane cooling component) to avoid exaggerating the related torque.

In the more realistic scenario of a planet migrating through the disk as it grows, we expect that the effect discussed in this work will be less important during the earlier stages of migration, when the planet is still small and the corotating region is not yet fully developed. However, as the planet grows, the corotating region widens, and the cooling-induced vortensity growth becomes stronger, its impact on migration will quickly become relevant. Finding a critical mass beyond which this happens is beyond the scope of our work, however.

Finally, as we have discussed, it becomes apparent that holding the planet in a fixed radial location would both hide the contribution of this effect to the total torque, as it mainly affects the dynamical corotation torque, but also artificially speed up inward migration once the planet is released due to the pileup of vortensity in the corotating region. For this reason, we recommend that numerical models of planet migration allow the planet to migrate as soon as possible after the start of the simulation.

\subsection{Connecting to population synthesis models}
\label{sub:relevance}

As discussed in Sect.~\ref{sec:introduction}, we expect that dynamical corotation torques will be important for low- to moderate-mass planets in low-turbulence disks. In order to investigate its effects on planet populations, however, a recipe for the dynamical corotation torque is required. \citet{mcnally-etal-2018} provided such a recipe for a planet migrating in a disk where the Hall effect drives a Lorentz acceleration on the disk, resulting in a background radial velocity $u_R$ and a vortensity source term on the planet given by
\begin{equation}
	\label{eq:mcnally-source}
	\D{\varpih}{t} = - \varpih^2 \left[\frac{1}{\Sigma R}\D{}{R}\left(R a_\phi\right)\right], \qquad a_\phi = \frac{\sqrt{\G\Mstar R}}{2R^2} u_R.
\end{equation}
The vortensity source term $\mathcal{S}$ and the associated torque that we discuss in this paper is not related to magnetic fields as in the work of \citet{mcnally-etal-2018}, but it would be useful to parametrize $\mathcal{S}$ as a function of planet and disk properties such that its effects can be included in population synthesis models. This will be the focus of followup work.

At the same time, it is important to note that our models focus on disks with a single planet. In a multiple-planet configuration, on top of planet--planet interactions, it is possible that the vortensity-rich streamlines left behind by the innermost, inwardly-migrating planet can interfere with the migration of planets further out by increasing the local disk vortensity above its nominal Keplerian value. This could enhance the dynamical corotation torque on the remaining planets, slowing them down.

\section{Summary}
\label{sec:summary}

We carried out numerical simulations of planet--disk interaction in inviscid disks, and investigated the effect of cooling on the vortensity evolution in the planet's corotating region. We summarize our findings below.

We found that radiative cooling interferes with the otherwise adiabatic compression and expansion of material in the planet's corotating region as it U-turns near the planet. The result is a net increase in vortensity in the corotating region, which weakens the stalling effect of the dynamical corotation torque \citep{mcnally-etal-2017} and substantially speeds up inward migration. We identified the reason for this vortensity growth as baroclinic forcing along streamlines U-turning very close to the planet, causing the vortensity excess to pile up at the edges of the corotating region.

We then compared models of migrating planets with and without radiative cooling, and found that this baroclinic forcing results in slightly faster inward migration for the first 200 orbits. As the planet continues to migrate inwards, a competition between the dynamical corotation torque becoming stronger for a rapidly-migrating planet and baroclinic forcing enhancing vortensity growth for a slowly-migrating planet prevents the planet from stalling and instead allows it to sustain a migration rate 2--3$\times$ faster than  the isothermal or adiabatic rates. We also found that the vortensity growth rate (and therefore inward migration rate) is maximized when the cooling timescale is comparable to the U-turn timescale of material in the corotating region, making this mechanism relevant in a quite broad radial range in the disk ($R\sim5\text{--}50$\,au).

Furthermore, we showed that thermal diffusion acts to suppress vortensity growth due to baroclinic forcing. We therefore conclude that the effect of radiative cooling on inward migration is twofold: it acts to speed up inward migration via vortensity growth, but this growth is partly suppressed by thermal diffusion.

Finally, we discussed how radiative damping of spiral shocks can further enhance the effect of cooling on inward migration by weakening the vortensity barrier created by the planet's own spiral shocks. While this effect is of secondary importance compared to the baroclinic forcing we discuss in this paper, we expect that it will be relevant in the radial range where $\beta\sim0.1\text{--}10$ \citep{miranda-rafikov-2020a,ziampras-etal-2023}, and note that the two effects could operate efficiently together for $\beta\sim1\text{--}10$.

Our results highlight the importance of radiative effects in planet--disk interaction, and in particular the role of thermal diffusion. An important take-home message is that type-I migration in low-turbulence, radiative disks might be more challenging to understand than previously thought, and mechanisms that can naturally lead to outward migration (e.g., magnetic fields, planetary accretion luminosity) will become necessary from a modeling perspective in order to explain the population of low-mass planets at $R\sim1\text{--}10$\,au. We also expect that the effect of thermal diffusion will be even more central when additional thermal effects are considered.

\section*{Acknowledgements}
AZ would like to thank Roman Rafikov, Kees Dullemond and Josh Brown for their suggestions and helpful discussions. This research utilized Queen Mary's Apocrita HPC facility, supported by QMUL Research-IT (http://doi.org/10.5281/zenodo.438045). This work was performed using the DiRAC Data Intensive service at Leicester, operated by the University of Leicester IT Services, which forms part of the STFC DiRAC HPC Facility (www.dirac.ac.uk). The equipment was funded by BEIS capital funding via STFC capital grants ST/K000373/1 and ST/R002363/1 and STFC DiRAC Operations grant ST/R001014/1. DiRAC is part of the National e-Infrastructure. AZ and RPN are supported by STFC grant ST/P000592/1, and RPN is supported by the Leverhulme Trust through grant RPG-2018-418. This project has received funding from the European Research Council (ERC) under the European Union's Horizon 2020 research and innovation programme (grant agreement No 101054502). All plots in this paper were made with the Python library \texttt{matplotlib} \citep{hunter-2007}.

\section*{Data Availability}

Data from our numerical models are available upon reasonable request to the corresponding author.

\bibliographystyle{mnras}
\bibliography{refs}

\begin{thebibliography}{}
\makeatletter
\relax
\def\mn@urlcharsother{\let\do\@makeother \do\$\do\&\do\#\do\^\do\_\do\%\do\~}
\def\mn@doi{\begingroup\mn@urlcharsother \@ifnextchar [ {\mn@doi@}
  {\mn@doi@[]}}
\def\mn@doi@[#1]#2{\def\@tempa{#1}\ifx\@tempa\@empty \href
  {http://dx.doi.org/#2} {doi:#2}\else \href {http://dx.doi.org/#2} {#1}\fi
  \endgroup}
\def\mn@eprint#1#2{\mn@eprint@#1:#2::\@nil}
\def\mn@eprint@arXiv#1{\href {http://arxiv.org/abs/#1} {{\tt arXiv:#1}}}
\def\mn@eprint@dblp#1{\href {http://dblp.uni-trier.de/rec/bibtex/#1.xml}
  {dblp:#1}}
\def\mn@eprint@#1:#2:#3:#4\@nil{\def\@tempa {#1}\def\@tempb {#2}\def\@tempc
  {#3}\ifx \@tempc \@empty \let \@tempc \@tempb \let \@tempb \@tempa \fi \ifx
  \@tempb \@empty \def\@tempb {arXiv}\fi \@ifundefined
  {mn@eprint@\@tempb}{\@tempb:\@tempc}{\expandafter \expandafter \csname
  mn@eprint@\@tempb\endcsname \expandafter{\@tempc}}}

\bibitem[\protect\citeauthoryear{{Andrews} et~al.,}{{Andrews}
  et~al.}{2018}]{andrews-etal-2018}
{Andrews} S.~M.,  et~al., 2018, \mn@doi [\apjl] {10.3847/2041-8213/aaf741},
  \href {https://ui.adsabs.harvard.edu/abs/2018ApJ...869L..41A} {869, L41}

\bibitem[\protect\citeauthoryear{{Bai} \& {Stone}}{{Bai} \&
  {Stone}}{2013}]{bai-stone-2013}
{Bai} X.-N.,  {Stone} J.~M.,  2013, \mn@doi [\apj]
  {10.1088/0004-637X/769/1/76}, \href
  {https://ui.adsabs.harvard.edu/abs/2013ApJ...769...76B} {769, 76}

\bibitem[\protect\citeauthoryear{{Baruteau} \& {Masset}}{{Baruteau} \&
  {Masset}}{2008a}]{baruteau-masset-2008a}
{Baruteau} C.,  {Masset} F.,  2008a, \mn@doi [\apj] {10.1086/523667}, \href
  {https://ui.adsabs.harvard.edu/abs/2008ApJ...672.1054B} {672, 1054}

\bibitem[\protect\citeauthoryear{{Baruteau} \& {Masset}}{{Baruteau} \&
  {Masset}}{2008b}]{baruteau-masset-2008b}
{Baruteau} C.,  {Masset} F.,  2008b, \mn@doi [\apj] {10.1086/529487}, \href
  {https://ui.adsabs.harvard.edu/abs/2008ApJ...678..483B} {678, 483}

\bibitem[\protect\citeauthoryear{{Ben{\'\i}tez-Llambay}, {Masset},
  {Koenigsberger}  \& {Szul{\'a}gyi}}{{Ben{\'\i}tez-Llambay}
  et~al.}{2015}]{benitez-etal-2015}
{Ben{\'\i}tez-Llambay} P.,  {Masset} F.,  {Koenigsberger} G.,   {Szul{\'a}gyi}
  J.,  2015, \mn@doi [\nat] {10.1038/nature14277}, \href
  {https://ui.adsabs.harvard.edu/abs/2015Natur.520...63B} {520, 63}

\bibitem[\protect\citeauthoryear{{Birnstiel} et~al.,}{{Birnstiel}
  et~al.}{2018}]{birnstiel-etal-2018}
{Birnstiel} T.,  et~al., 2018, \mn@doi [The Astrophysical Journal]
  {10.3847/2041-8213/aaf743}, \href
  {https://ui.adsabs.harvard.edu/abs/2018ApJ...869L..45B} {869, L45}

\bibitem[\protect\citeauthoryear{{Casoli} \& {Masset}}{{Casoli} \&
  {Masset}}{2009}]{casoli-masset-2009}
{Casoli} J.,  {Masset} F.~S.,  2009, \mn@doi [\apj]
  {10.1088/0004-637X/703/1/845}, \href
  {https://ui.adsabs.harvard.edu/abs/2009ApJ...703..845C} {703, 845}

\bibitem[\protect\citeauthoryear{{Chiang} \& {Goldreich}}{{Chiang} \&
  {Goldreich}}{1997}]{chiang-goldreich-1997}
{Chiang} E.~I.,  {Goldreich} P.,  1997, \mn@doi [\apj] {10.1086/304869}, \href
  {https://ui.adsabs.harvard.edu/abs/1997ApJ...490..368C} {490, 368}

\bibitem[\protect\citeauthoryear{{Coleman} \& {Nelson}}{{Coleman} \&
  {Nelson}}{2014}]{coleman-nelson-2014}
{Coleman} G. A.~L.,  {Nelson} R.~P.,  2014, \mn@doi [\mnras]
  {10.1093/mnras/stu1715}, \href
  {https://ui.adsabs.harvard.edu/abs/2014MNRAS.445..479C} {445, 479}

\bibitem[\protect\citeauthoryear{{Coleman} \& {Nelson}}{{Coleman} \&
  {Nelson}}{2016}]{coleman-nelson-2016}
{Coleman} G. A.~L.,  {Nelson} R.~P.,  2016, \mn@doi [\mnras]
  {10.1093/mnras/stw1177}, \href
  {https://ui.adsabs.harvard.edu/abs/2016MNRAS.460.2779C} {460, 2779}

\bibitem[\protect\citeauthoryear{{Crida}, {Morbidelli}  \& {Masset}}{{Crida}
  et~al.}{2006}]{crida-etal-2006}
{Crida} A.,  {Morbidelli} A.,   {Masset} F.,  2006, \mn@doi [\icarus]
  {10.1016/j.icarus.2005.10.007}, \href
  {https://ui.adsabs.harvard.edu/abs/2006Icar..181..587C} {181, 587}

\bibitem[\protect\citeauthoryear{{Crida}, {Griveaud}, {Lega}, {Masset},
  {Morbidelli}, {Kloster}, {Marques}  \& {Minker}}{{Crida}
  et~al.}{2022}]{crida-etal-2022}
{Crida} A.,  {Griveaud} P.,  {Lega} E.,  {Masset} F.,  {Morbidelli} A.,
  {Kloster} D.,  {Marques} L.,   {Minker} K.,  2022, in {Richard} J.,  et~al.,
  eds, SF2A-2022: Proceedings of the Annual meeting of the French Society of
  Astronomy and Astrophysics. Eds.: J. Richard. pp 315--317

\bibitem[\protect\citeauthoryear{{Cummins}, {Owen}  \& {Booth}}{{Cummins}
  et~al.}{2022}]{cummins-etal-2022}
{Cummins} D.~P.,  {Owen} J.~E.,   {Booth} R.~A.,  2022, \mn@doi [\mnras]
  {10.1093/mnras/stac1819}, \href
  {https://ui.adsabs.harvard.edu/abs/2022MNRAS.515.1276C} {515, 1276}

\bibitem[\protect\citeauthoryear{{Dullemond}, {Ziampras}, {Ostertag}  \&
  {Dominik}}{{Dullemond} et~al.}{2022}]{dullemond-etal-2022}
{Dullemond} C.~P.,  {Ziampras} A.,  {Ostertag} D.,   {Dominik} C.,  2022,
  \mn@doi [\aap] {10.1051/0004-6361/202244218}, \href
  {https://ui.adsabs.harvard.edu/abs/2022A&A...668A.105D} {668, A105}

\bibitem[\protect\citeauthoryear{Gammie}{Gammie}{2001}]{gammie-2001}
Gammie C.~F.,  2001, \mn@doi [\apj] {10.1086/320631}, 553, 174

\bibitem[\protect\citeauthoryear{{Goldreich} \& {Tremaine}}{{Goldreich} \&
  {Tremaine}}{1979}]{goldreich-tremaine-1979}
{Goldreich} P.,  {Tremaine} S.,  1979, \mn@doi [\apj] {10.1086/157448}, \href
  {https://ui.adsabs.harvard.edu/abs/1979ApJ...233..857G} {233, 857}

\bibitem[\protect\citeauthoryear{{Goodman} \& {Rafikov}}{{Goodman} \&
  {Rafikov}}{2001}]{goodman-rafikov-2001}
{Goodman} J.,  {Rafikov} R.~R.,  2001, \mn@doi [\apj] {10.1086/320572}, \href
  {https://ui.adsabs.harvard.edu/abs/2001ApJ...552..793G} {552, 793}

\bibitem[\protect\citeauthoryear{{Haffert}, {Bohn}, {de Boer}, {Snellen},
  {Brinchmann}, {Girard}, {Keller}  \& {Bacon}}{{Haffert}
  et~al.}{2019}]{haffert-etal-2019}
{Haffert} S.~Y.,  {Bohn} A.~J.,  {de Boer} J.,  {Snellen} I.~A.~G.,
  {Brinchmann} J.,  {Girard} J.~H.,  {Keller} C.~U.,   {Bacon} R.,  2019,
  \mn@doi [Nature Astronomy] {10.1038/s41550-019-0780-5}, \href
  {https://ui.adsabs.harvard.edu/abs/2019NatAs...3..749H} {3, 749}

\bibitem[\protect\citeauthoryear{{Hubeny}}{{Hubeny}}{1990}]{hubeny-1990}
{Hubeny} I.,  1990, \mn@doi [\apj] {10.1086/168501}, \href
  {http://adsabs.harvard.edu/abs/1990ApJ...351..632H} {351, 632}

\bibitem[\protect\citeauthoryear{Hunter}{Hunter}{2007}]{hunter-2007}
Hunter J.~D.,  2007, Computing In Science \& Engineering, 9, 90

\bibitem[\protect\citeauthoryear{{Izidoro}, {Ogihara}, {Raymond}, {Morbidelli},
  {Pierens}, {Bitsch}, {Cossou}  \& {Hersant}}{{Izidoro}
  et~al.}{2017}]{izidoro-etal-2017}
{Izidoro} A.,  {Ogihara} M.,  {Raymond} S.~N.,  {Morbidelli} A.,  {Pierens} A.,
   {Bitsch} B.,  {Cossou} C.,   {Hersant} F.,  2017, \mn@doi [\mnras]
  {10.1093/mnras/stx1232}, \href
  {https://ui.adsabs.harvard.edu/abs/2017MNRAS.470.1750I} {470, 1750}

\bibitem[\protect\citeauthoryear{{Keppler} et~al.,}{{Keppler}
  et~al.}{2018}]{keppler-etal-2018}
{Keppler} M.,  et~al., 2018, \mn@doi [\aap] {10.1051/0004-6361/201832957},
  \href {https://ui.adsabs.harvard.edu/abs/2018A&A...617A..44K} {617, A44}

\bibitem[\protect\citeauthoryear{{Kimmig}, {Dullemond}  \& {Kley}}{{Kimmig}
  et~al.}{2020}]{kimmig-etal-2020}
{Kimmig} C.~N.,  {Dullemond} C.~P.,   {Kley} W.,  2020, \mn@doi [\aap]
  {10.1051/0004-6361/201936412}, \href
  {https://ui.adsabs.harvard.edu/abs/2020A&A...633A...4K} {633, A4}

\bibitem[\protect\citeauthoryear{{Kley}}{{Kley}}{1989}]{kley-1989}
{Kley} W.,  1989, \aap, \href
  {https://ui.adsabs.harvard.edu/abs/1989A&A...208...98K} {208, 98}

\bibitem[\protect\citeauthoryear{{Kley} \& {Crida}}{{Kley} \&
  {Crida}}{2008}]{kley-crida-2008}
{Kley} W.,  {Crida} A.,  2008, \mn@doi [\aap] {10.1051/0004-6361:200810033},
  \href {https://ui.adsabs.harvard.edu/abs/2008A&A...487L...9K} {487, L9}

\bibitem[\protect\citeauthoryear{{Kley} \& {Nelson}}{{Kley} \&
  {Nelson}}{2012}]{kley-nelson-2012}
{Kley} W.,  {Nelson} R.~P.,  2012, \mn@doi [\araa]
  {10.1146/annurev-astro-081811-125523}, \href
  {https://ui.adsabs.harvard.edu/abs/2012ARA&A..50..211K} {50, 211}

\bibitem[\protect\citeauthoryear{{Lega}, {Crida}, {Bitsch}  \&
  {Morbidelli}}{{Lega} et~al.}{2014}]{lega-etal-2014}
{Lega} E.,  {Crida} A.,  {Bitsch} B.,   {Morbidelli} A.,  2014, \mn@doi
  [\mnras] {10.1093/mnras/stu304}, \href
  {https://ui.adsabs.harvard.edu/abs/2014MNRAS.440..683L} {440, 683}

\bibitem[\protect\citeauthoryear{{Lega} et~al.,}{{Lega}
  et~al.}{2021}]{lega-etal-2021}
{Lega} E.,  et~al., 2021, \mn@doi [\aap] {10.1051/0004-6361/202039520}, \href
  {https://ui.adsabs.harvard.edu/abs/2021A&A...646A.166L} {646, A166}

\bibitem[\protect\citeauthoryear{{Levermore} \& {Pomraning}}{{Levermore} \&
  {Pomraning}}{1981}]{levermore-pomraning-1981}
{Levermore} C.~D.,  {Pomraning} G.~C.,  1981, \mn@doi [\apj] {10.1086/159157},
  \href {https://ui.adsabs.harvard.edu/abs/1981ApJ...248..321L} {248, 321}

\bibitem[\protect\citeauthoryear{{Lin} \& {Papaloizou}}{{Lin} \&
  {Papaloizou}}{1985}]{lin-papaloizou-1985}
{Lin} D.~N.~C.,  {Papaloizou} J.,  1985, in {Black} D.~C.,  {Matthews} M.~S.,
  eds, Protostars and Planets II. pp 981--1072

\bibitem[\protect\citeauthoryear{{Lin} \& {Papaloizou}}{{Lin} \&
  {Papaloizou}}{2010}]{lin-papaloizou-2010}
{Lin} M.-K.,  {Papaloizou} J. C.~B.,  2010, \mn@doi [\mnras]
  {10.1111/j.1365-2966.2010.16560.x}, \href
  {https://ui.adsabs.harvard.edu/abs/2010MNRAS.405.1473L} {405, 1473}

\bibitem[\protect\citeauthoryear{Lovelace, Li, Colgate  \& Nelson}{Lovelace
  et~al.}{1999}]{lovelace-1999}
Lovelace R. V.~E.,  Li H.,  Colgate S.~A.,   Nelson A.~F.,  1999, \mn@doi
  [\apj] {10.1086/306900}, 513, 805

\bibitem[\protect\citeauthoryear{{Masset}}{{Masset}}{2000}]{masset-2000}
{Masset} F.,  2000, \mn@doi [\aaps] {10.1051/aas:2000116}, \href
  {http://adsabs.harvard.edu/abs/2000A%26AS..141..165M} {141, 165}

\bibitem[\protect\citeauthoryear{{Masset}}{{Masset}}{2017}]{masset-2017}
{Masset} F.~S.,  2017, \mn@doi [\mnras] {10.1093/mnras/stx2271}, \href
  {https://ui.adsabs.harvard.edu/abs/2017MNRAS.472.4204M} {472, 4204}

\bibitem[\protect\citeauthoryear{{Masset} \& {Papaloizou}}{{Masset} \&
  {Papaloizou}}{2003}]{masset-papaloizou-2003}
{Masset} F.~S.,  {Papaloizou} J.~C.~B.,  2003, \mn@doi [\apj] {10.1086/373892},
  \href {https://ui.adsabs.harvard.edu/abs/2003ApJ...588..494M} {588, 494}

\bibitem[\protect\citeauthoryear{{McNally}, {Nelson}, {Paardekooper}, {Gressel}
   \& {Lyra}}{{McNally} et~al.}{2017}]{mcnally-etal-2017}
{McNally} C.~P.,  {Nelson} R.~P.,  {Paardekooper} S.-J.,  {Gressel} O.,
  {Lyra} W.,  2017, \mn@doi [\mnras] {10.1093/mnras/stx2136}, \href
  {https://ui.adsabs.harvard.edu/abs/2017MNRAS.472.1565M} {472, 1565}

\bibitem[\protect\citeauthoryear{{McNally}, {Nelson}  \&
  {Paardekooper}}{{McNally} et~al.}{2018}]{mcnally-etal-2018}
{McNally} C.~P.,  {Nelson} R.~P.,   {Paardekooper} S.-J.,  2018, \mn@doi
  [\mnras] {10.1093/mnras/sty905}, \href
  {https://ui.adsabs.harvard.edu/abs/2018MNRAS.477.4596M} {477, 4596}

\bibitem[\protect\citeauthoryear{{McNally}, {Nelson}, {Paardekooper}  \&
  {Ben{\'\i}tez-Llambay}}{{McNally} et~al.}{2019}]{mcnally-etal-2019a}
{McNally} C.~P.,  {Nelson} R.~P.,  {Paardekooper} S.-J.,
  {Ben{\'\i}tez-Llambay} P.,  2019, \mn@doi [\mnras] {10.1093/mnras/stz023},
  \href {https://ui.adsabs.harvard.edu/abs/2019MNRAS.484..728M} {484, 728}

\bibitem[\protect\citeauthoryear{{McNally}, {Nelson}, {Paardekooper},
  {Ben{\'\i}tez-Llambay}  \& {Gressel}}{{McNally}
  et~al.}{2020}]{mcnally-etal-2020}
{McNally} C.~P.,  {Nelson} R.~P.,  {Paardekooper} S.-J.,
  {Ben{\'\i}tez-Llambay} P.,   {Gressel} O.,  2020, \mn@doi [\mnras]
  {10.1093/mnras/staa576}, \href
  {https://ui.adsabs.harvard.edu/abs/2020MNRAS.493.4382M} {493, 4382}

\bibitem[\protect\citeauthoryear{{Menou} \& {Goodman}}{{Menou} \&
  {Goodman}}{2004}]{menou-goodman-2004}
{Menou} K.,  {Goodman} J.,  2004, \mn@doi [\apj] {10.1086/382947}, \href
  {https://ui.adsabs.harvard.edu/abs/2004ApJ...606..520M} {606, 520}

\bibitem[\protect\citeauthoryear{{Mignone}, {Bodo}, {Massaglia}, {Matsakos},
  {Tesileanu}, {Zanni}  \& {Ferrari}}{{Mignone}
  et~al.}{2007}]{mignone-etal-2007}
{Mignone} A.,  {Bodo} G.,  {Massaglia} S.,  {Matsakos} T.,  {Tesileanu} O.,
  {Zanni} C.,   {Ferrari} A.,  2007, \mn@doi [The Astrophysical Journal
  Supplement Series] {10.1086/513316}, \href
  {https://ui.adsabs.harvard.edu/\#abs/2007ApJS..170..228M} {170, 228}

\bibitem[\protect\citeauthoryear{{Mignone}, {Flock}, {Stute}, {Kolb}  \&
  {Muscianisi}}{{Mignone} et~al.}{2012}]{mignone-etal-2012}
{Mignone} A.,  {Flock} M.,  {Stute} M.,  {Kolb} S.~M.,   {Muscianisi} G.,
  2012, \mn@doi [\aap] {10.1051/0004-6361/201219557}, \href
  {http://adsabs.harvard.edu/abs/2012A%26A...545A.152M} {545, A152}

\bibitem[\protect\citeauthoryear{{Miranda} \& {Rafikov}}{{Miranda} \&
  {Rafikov}}{2019}]{miranda-rafikov-2019}
{Miranda} R.,  {Rafikov} R.~R.,  2019, \mn@doi [\apjl]
  {10.3847/2041-8213/ab22a7}, \href
  {https://ui.adsabs.harvard.edu/abs/2019ApJ...878L...9M} {878, L9}

\bibitem[\protect\citeauthoryear{{Miranda} \& {Rafikov}}{{Miranda} \&
  {Rafikov}}{2020a}]{miranda-rafikov-2020a}
{Miranda} R.,  {Rafikov} R.~R.,  2020a, \mn@doi [\apj]
  {10.3847/1538-4357/ab791a}, \href
  {https://ui.adsabs.harvard.edu/abs/2020ApJ...892...65M} {892, 65}

\bibitem[\protect\citeauthoryear{{Miranda} \& {Rafikov}}{{Miranda} \&
  {Rafikov}}{2020b}]{miranda-rafikov-2020b}
{Miranda} R.,  {Rafikov} R.~R.,  2020b, \mn@doi [\apj]
  {10.3847/1538-4357/abbee7}, \href
  {https://ui.adsabs.harvard.edu/abs/2020ApJ...904..121M} {904, 121}

\bibitem[\protect\citeauthoryear{{Mordasini}, {Alibert}  \& {Benz}}{{Mordasini}
  et~al.}{2009a}]{mordasini-etal-2009a}
{Mordasini} C.,  {Alibert} Y.,   {Benz} W.,  2009a, \mn@doi [\aap]
  {10.1051/0004-6361/200810301}, \href
  {https://ui.adsabs.harvard.edu/abs/2009A&A...501.1139M} {501, 1139}

\bibitem[\protect\citeauthoryear{{Mordasini}, {Alibert}, {Benz}  \&
  {Naef}}{{Mordasini} et~al.}{2009b}]{mordasini-etal-2009b}
{Mordasini} C.,  {Alibert} Y.,  {Benz} W.,   {Naef} D.,  2009b, \mn@doi [\aap]
  {10.1051/0004-6361/200810697}, \href
  {https://ui.adsabs.harvard.edu/abs/2009A&A...501.1161M} {501, 1161}

\bibitem[\protect\citeauthoryear{{M{\"u}ller} \& {Kley}}{{M{\"u}ller} \&
  {Kley}}{2012}]{mueller-kley-2012}
{M{\"u}ller} T.~W.~A.,  {Kley} W.,  2012, \mn@doi [\aap]
  {10.1051/0004-6361/201118202}, \href
  {https://ui.adsabs.harvard.edu/abs/2012A&A...539A..18M} {539, A18}

\bibitem[\protect\citeauthoryear{{{\"O}berg} et~al.,}{{{\"O}berg}
  et~al.}{2021}]{oberg-etal-2021}
{{\"O}berg} K.~I.,  et~al., 2021, \mn@doi [\apjs] {10.3847/1538-4365/ac1432},
  \href {https://ui.adsabs.harvard.edu/abs/2021ApJS..257....1O} {257, 1}

\bibitem[\protect\citeauthoryear{{Paardekooper}}{{Paardekooper}}{2014}]{paardekooper-2014}
{Paardekooper} S.~J.,  2014, \mn@doi [\mnras] {10.1093/mnras/stu1542}, \href
  {https://ui.adsabs.harvard.edu/abs/2014MNRAS.444.2031P} {444, 2031}

\bibitem[\protect\citeauthoryear{{Paardekooper}, {Lesur}  \&
  {Papaloizou}}{{Paardekooper} et~al.}{2010}]{paardekooper-etal-2010}
{Paardekooper} S.-J.,  {Lesur} G.,   {Papaloizou} J. C.~B.,  2010, \mn@doi
  [\apj] {10.1088/0004-637X/725/1/146}, \href
  {https://ui.adsabs.harvard.edu/abs/2010ApJ...725..146P} {725, 146}

\bibitem[\protect\citeauthoryear{{Paardekooper}, {Baruteau}  \&
  {Kley}}{{Paardekooper} et~al.}{2011}]{paardekooper-etal-2011}
{Paardekooper} S.~J.,  {Baruteau} C.,   {Kley} W.,  2011, \mn@doi [\mnras]
  {10.1111/j.1365-2966.2010.17442.x}, \href
  {https://ui.adsabs.harvard.edu/abs/2011MNRAS.410..293P} {410, 293}

\bibitem[\protect\citeauthoryear{{Paardekooper}, {Dong}, {Duffell}, {Fung},
  {Masset}, {Ogilvie}  \& {Tanaka}}{{Paardekooper}
  et~al.}{2022}]{paardekooper-etal-2022}
{Paardekooper} S.-J.,  {Dong} R.,  {Duffell} P.,  {Fung} J.,  {Masset} F.~S.,
  {Ogilvie} G.,   {Tanaka} H.,  2022, \mn@doi [arXiv e-prints]
  {10.48550/arXiv.2203.09595}, \href
  {https://ui.adsabs.harvard.edu/abs/2022arXiv220309595P} {p. arXiv:2203.09595}

\bibitem[\protect\citeauthoryear{{Papaloizou}, {Nelson}, {Kley}, {Masset}  \&
  {Artymowicz}}{{Papaloizou} et~al.}{2007}]{papaloizou-etal-2007}
{Papaloizou} J.~C.~B.,  {Nelson} R.~P.,  {Kley} W.,  {Masset} F.~S.,
  {Artymowicz} P.,  2007, in {Reipurth} B.,  {Jewitt} D.,   {Keil} K.,  eds,
  Protostars and Planets V. p.~655 (\mn@eprint {arXiv} {astro-ph/0603196}),
  \mn@doi{10.48550/arXiv.astro-ph/0603196}

\bibitem[\protect\citeauthoryear{{Pierens}}{{Pierens}}{2015}]{pierens-2015}
{Pierens} A.,  2015, \mn@doi [\mnras] {10.1093/mnras/stv2024}, \href
  {https://ui.adsabs.harvard.edu/abs/2015MNRAS.454.2003P} {454, 2003}

\bibitem[\protect\citeauthoryear{{Rafikov}}{{Rafikov}}{2002}]{rafikov-2002}
{Rafikov} R.~R.,  2002, \mn@doi [\apj] {10.1086/339399}, \href
  {https://ui.adsabs.harvard.edu/abs/2002ApJ...569..997R} {569, 997}

\bibitem[\protect\citeauthoryear{{Tanaka}, {Takeuchi}  \& {Ward}}{{Tanaka}
  et~al.}{2002}]{tanaka-etal-2002}
{Tanaka} H.,  {Takeuchi} T.,   {Ward} W.~R.,  2002, \mn@doi [\apj]
  {10.1086/324713}, \href
  {https://ui.adsabs.harvard.edu/abs/2002ApJ...565.1257T} {565, 1257}

\bibitem[\protect\citeauthoryear{{Teague}, {Bae}, {Bergin}, {Birnstiel}  \&
  {Foreman-Mackey}}{{Teague} et~al.}{2018}]{teague-etal-2018}
{Teague} R.,  {Bae} J.,  {Bergin} E.~A.,  {Birnstiel} T.,   {Foreman-Mackey}
  D.,  2018, \mn@doi [\apjl] {10.3847/2041-8213/aac6d7}, \href
  {https://ui.adsabs.harvard.edu/abs/2018ApJ...860L..12T} {860, L12}

\bibitem[\protect\citeauthoryear{{Thun} \& {Kley}}{{Thun} \&
  {Kley}}{2018}]{thun-kley-2018}
{Thun} D.,  {Kley} W.,  2018, \mn@doi [\aap] {10.1051/0004-6361/201832804},
  \href {https://ui.adsabs.harvard.edu/abs/2018A&A...616A..47T} {616, A47}

\bibitem[\protect\citeauthoryear{Toro, Spruce  \& Speares}{Toro
  et~al.}{1994}]{toro-etal-1994}
Toro E.~F.,  Spruce M.,   Speares W.,  1994, Shock waves, 4, 25

\bibitem[\protect\citeauthoryear{Van~Leer}{Van~Leer}{1974}]{vanleer-1974}
Van~Leer B.,  1974, Journal of computational physics, 14, 361

\bibitem[\protect\citeauthoryear{{Wafflard-Fernandez} \&
  {Lesur}}{{Wafflard-Fernandez} \&
  {Lesur}}{2023}]{wafflard-fernandez-lesur-2023}
{Wafflard-Fernandez} G.,  {Lesur} G.,  2023, \mn@doi [arXiv e-prints]
  {10.48550/arXiv.2305.11784}, \href
  {https://ui.adsabs.harvard.edu/abs/2023arXiv230511784W} {p. arXiv:2305.11784}

\bibitem[\protect\citeauthoryear{{Ward}}{{Ward}}{1997a}]{ward-1997a}
{Ward} W.~R.,  1997a, \mn@doi [\icarus] {10.1006/icar.1996.5647}, \href
  {https://ui.adsabs.harvard.edu/abs/1997Icar..126..261W} {126, 261}

\bibitem[\protect\citeauthoryear{{Ward}}{{Ward}}{1997b}]{ward-1997b}
{Ward} W.~R.,  1997b, \mn@doi [\apjl] {10.1086/310701}, \href
  {https://ui.adsabs.harvard.edu/abs/1997ApJ...482L.211W} {482, L211}

\bibitem[\protect\citeauthoryear{{Yamaleev} \& {Carpenter}}{{Yamaleev} \&
  {Carpenter}}{2009}]{yamaleev-carpenter-2009}
{Yamaleev} N.~K.,  {Carpenter} M.~H.,  2009, \mn@doi [Journal of Computational
  Physics] {10.1016/j.jcp.2009.01.011}, \href
  {https://ui.adsabs.harvard.edu/abs/2009JCoPh.228.3025Y} {228, 3025}

\bibitem[\protect\citeauthoryear{{Yun}, {Kim}, {Bae}  \& {Han}}{{Yun}
  et~al.}{2022}]{yun-etal-2022}
{Yun} H.-G.,  {Kim} W.-T.,  {Bae} J.,   {Han} C.,  2022, \mn@doi [\apj]
  {10.3847/1538-4357/ac9185}, \href
  {https://ui.adsabs.harvard.edu/abs/2022ApJ...938..102Y} {938, 102}

\bibitem[\protect\citeauthoryear{{Zhang} \& {Zhu}}{{Zhang} \&
  {Zhu}}{2020}]{zhang-zhu-2020}
{Zhang} S.,  {Zhu} Z.,  2020, \mn@doi [\mnras] {10.1093/mnras/staa404}, \href
  {https://ui.adsabs.harvard.edu/abs/2020MNRAS.493.2287Z} {493, 2287}

\bibitem[\protect\citeauthoryear{{Zhu}, {Stone}  \& {Rafikov}}{{Zhu}
  et~al.}{2012}]{zhu-etal-2012}
{Zhu} Z.,  {Stone} J.~M.,   {Rafikov} R.~R.,  2012, \mn@doi [\apjl]
  {10.1088/2041-8205/758/2/L42}, \href
  {https://ui.adsabs.harvard.edu/abs/2012ApJ...758L..42Z} {758, L42}

\bibitem[\protect\citeauthoryear{{Zhu}, {Dong}, {Stone}  \& {Rafikov}}{{Zhu}
  et~al.}{2015}]{zhu-etal-2015}
{Zhu} Z.,  {Dong} R.,  {Stone} J.~M.,   {Rafikov} R.~R.,  2015, \mn@doi [\apj]
  {10.1088/0004-637X/813/2/88}, \href
  {https://ui.adsabs.harvard.edu/abs/2015ApJ...813...88Z} {813, 88}

\bibitem[\protect\citeauthoryear{{Ziampras}, {Kley}  \& {Dullemond}}{{Ziampras}
  et~al.}{2020}]{ziampras-etal-2020b}
{Ziampras} A.,  {Kley} W.,   {Dullemond} C.~P.,  2020, \mn@doi [\aap]
  {10.1051/0004-6361/201937048}, \href
  {https://ui.adsabs.harvard.edu/abs/2020A&A...637A..50Z} {637, A50}

\bibitem[\protect\citeauthoryear{{Ziampras}, {Nelson}  \& {Rafikov}}{{Ziampras}
  et~al.}{2023}]{ziampras-etal-2023}
{Ziampras} A.,  {Nelson} R.~P.,   {Rafikov} R.~R.,  2023, \mn@doi [arXiv
  e-prints] {10.48550/arXiv.2305.14415}, \href
  {https://ui.adsabs.harvard.edu/abs/2023arXiv230514415Z} {p. arXiv:2305.14415}

\bibitem[\protect\citeauthoryear{{de Val-Borro} et~al.,}{{de Val-Borro}
  et~al.}{2006}]{devalborro-etal-2006}
{de Val-Borro} M.,  et~al., 2006, \mn@doi [\mnras]
  {10.1111/j.1365-2966.2006.10488.x}, \href
  {http://adsabs.harvard.edu/abs/2006MNRAS.370..529D} {370, 529}

\makeatother
\end{thebibliography}

\appendix

\section{Vortensity growth with cooling}
\label{apdx:vortensity-math}

We follow a similar approach to Sect.~\ref{sub:vortensity-growth}, where we look at a gas parcel U-turning ahead of or behind the planet. We can rewrite the source term $\mathcal{S}$ in Eq.~\eqref{eq:vortensity-equation} using the ideal gas law as
\begin{equation}
	\label{eq:vortensity-source-full}
	\mathcal{S} = \frac{P}{T\Sigma^3} \nabla\Sigma\times\nabla T = \frac{P}{T\Sigma^3} \left(\DP{\Sigma}{R}\DP{T}{\phi} - \DP{\Sigma}{\phi}\DP{T}{R}\right).
\end{equation}
Similar to Sect.~\ref{sub:vortensity-growth}, we now assume a hot and dense circular envelope around the planet such that $\partial_\phi\Sigma|_\text{b} = -\partial_\phi\Sigma|_\text{a} = \sigma_\phi$ and $\partial_\phi T|_\text{b} = -\partial_\phi T|_\text{a} = \tau_\phi$. We also assume that the gas parcel in a disk with cooling contracts and heats up in the same way that it would in an adiabatic disk, except that the expansion and cooling phases are different. In other words, $\sigma_\phi^\text{adb} = \sigma_\phi^\text{cool}=\sigma_\phi$ and $\tau_\phi^\text{adb} = \tau_\phi^\text{cool}=\tau_\phi$, where the superscripts denote the adiabatic and cooling cases.

In the adiabatic limit, $\mathcal{S}=0$ and therefore
\begin{equation}
	\mathcal{S}_\text{a}^\text{adb} = \frac{P_\text{a}}{T_\text{a}\Sigma_\text{a}^3}\left(-\left.\DP{\Sigma}{R}\right|_\text{a}^\text{adb}\tau_\phi+\left.\DP{T}{R}\right|_\text{a}^\text{adb} \sigma_\phi\right) = 0,
\end{equation}

\begin{equation}
	\mathcal{S}_\text{b}^\text{adb} = \frac{P_\text{b}}{T_\text{b}\Sigma_\text{b}^3}\left(\left.\DP{\Sigma}{R}\right|_\text{b}^\text{adb}\tau_\phi-\left.\DP{T}{R}\right|_\text{b}^\text{adb} \sigma_\phi\right) = 0.
\end{equation}

In the case with cooling, we assume that thermal emission and in-plane cooling happen alongside adiabatic expansion. The heated gas parcel will then cool faster (and expand less) than it would have adiabatically. We then write that
\begin{equation}
	\left.\DP{T}{R}\right|^\text{cool} \approx \left.\DP{T}{R}\right|^\text{adb} - \frac{|\delta T|}{\delta R}, \qquad \DP{\Sigma}{R} \approx \left.\DP{\Sigma}{R}\right|^\text{adb} + \frac{|\delta \Sigma|}{\delta R}.
\end{equation}
We then obtain ahead of the planet ($\delta R<0$)
\begin{equation}
    \label{eq:S-a-cool}
	\begin{split}
	\mathcal{S}_\text{a}^\text{cool} &\approx \frac{P_\text{a}}{T_\text{a}\Sigma_\text{a}^3}\left(-\left.\DP{\Sigma}{R}\right|_\text{a}^\text{cool}\tau_\phi+\left.\DP{T}{R}\right|_\text{a}^\text{cool} \sigma_\phi\right) \\
	&\approx \frac{1}{|\delta R|}\frac{P_\text{a}}{T_\text{a}\Sigma_\text{a}^3}\left(|\delta\Sigma|\,\tau_\phi + |\delta T|\,\sigma_\phi\right),
	\end{split}
\end{equation}
and behind the planet ($\delta R>0$)
\begin{equation}
    \label{eq:S-b-cool}
	\mathcal{S}_\text{b}^\text{cool} \approx \frac{1}{|\delta R|}\frac{P_\text{b}}{T_\text{b}\Sigma_\text{b}^3}\left(|\delta\Sigma|\,\tau_\phi + |\delta T|\,\sigma_\phi\right).
\end{equation}
As we have defined $\tau_\phi$ and $\sigma_\phi$ to be positive, we find that $\mathcal{S}^\text{cool} > 0$ during a U-turn both ahead of and behind the planet. In other words, vortensity will always increase along streamlines during a U-turn.

We note, however, that this is a crude approach as, in principle, the assumption that $\sigma_\phi^\text{adb}=\sigma_\phi^\text{cool}$ and $\tau_\phi^\text{adb}=\tau_\phi^\text{cool}$ will not hold. We also did not take into account the presence of a background temperature and density gradient, which would further complicate this calculation. We can nevertheless expect that, in any case, the two terms in Eqs.~\eqref{eq:S-a-cool}~\&~\eqref{eq:S-b-cool} will sum to a positive value such that there is a net vortensity growth.

\section{Vortensity growth in adiabatic models with a radial entropy gradient}
\label{apdx:adiabatic}

In Sect.~\ref{subsub:adiabatic}, we argued that a radial entropy gradient $K\propto R^\xi$ should result in a vortensity source term about the planet's azimuthal location. This term should however average to zero for each pair of streamlines U-turning ahead of and behind the planet, and only exist until the horseshoe is completely phase-mixed to a constant entropy state.
	
\begin{figure}
	\includegraphics[width=\columnwidth]{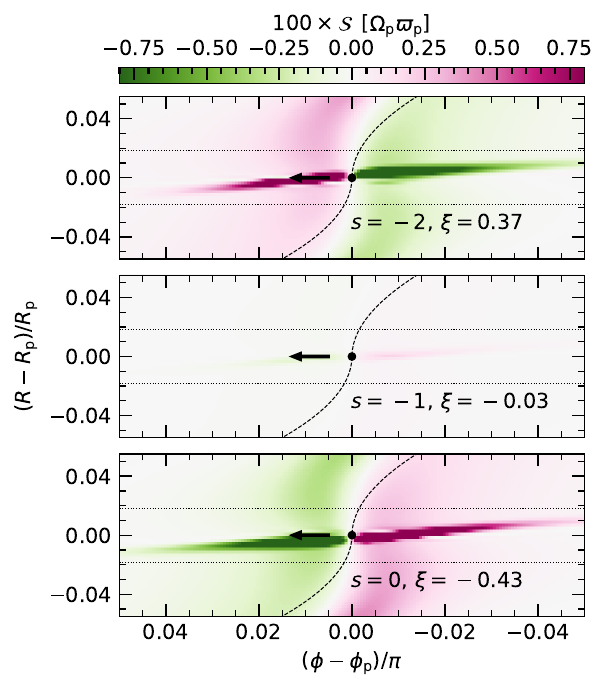}
	\caption{The baroclinic forcing term $\mathcal{S}$ in Eq.~\eqref{eq:vortensity-equation}, similar to Fig.~\ref{fig:fiducial-source}, for adiabatic models with different radial entropy exponents $\xi$, controlled by the surface density exponent $s$.}
	\label{fig:adiabatic-source}
\end{figure}

In Fig.~\ref{fig:adiabatic-source} we show the vortensity source term similar to Fig.~\ref{fig:fiducial-source} at $t=20$\,orbits for three adiabatic models with different surface density radial profiles $\Sigma\propto R^s$, where $s\in\{-2, -1, 0\}$. We find that the source term is zero throughout the corotating region for our fiducial model with $s=-1$ and $\xi\approx0$, but nonzero for the models with $\xi \neq 0$, where the source term is antisymmetric about the planet such that it averages to zero. This is consistent with our expectations in Sect.~\ref{subsub:adiabatic}.

\begin{figure}
	\includegraphics[width=\columnwidth]{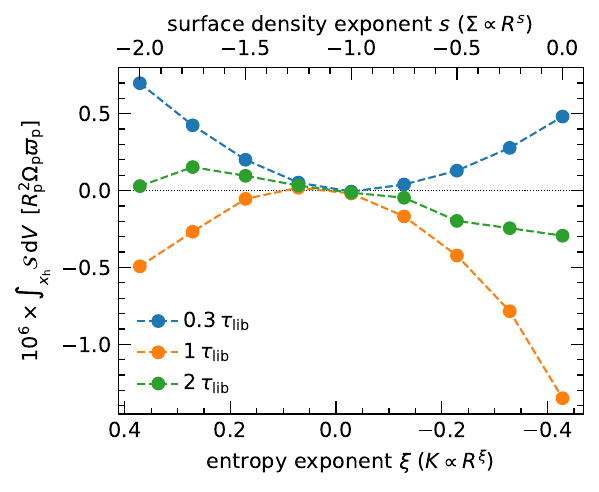}
	\caption{The baroclinic source term $\mathcal{S}$ in Eq.~\eqref{eq:vortensity-equation}, integrated over the corotating region similar to Fig.~\ref{fig:source-rad}, as a function of the radial entropy exponent $\xi$ and time. This term is very small compared to radiative models and decreases with time, consistent with our expectations in Sect.~\ref{subsub:adiabatic}.}
	\label{fig:source-adb}
\end{figure}

We further run a set of models with different surface density exponents $s$ to investigate the long-term evolution of $\mathcal{S}^\mathrm{adb}$ for different values of $\xi$. We compute the integrated vortensity source term inside the horseshoe similar to Sect.~\ref{sub:beta-dependency} and Fig.~\ref{fig:source-rad}, and plot the results as a function of $s$ and time in Fig.~\ref{fig:source-adb}. In line with our expectations, the source term is very small (about 100$\times$ smaller than a typical value in a radiative model) and further decreases with time.

\bsp	
\label{lastpage}
\end{document}